\documentclass[apj,twocolumn]{openjournal}
\usepackage{amsmath}
\usepackage{booktabs}
\usepackage{multirow}
\usepackage{color}
\usepackage{soul}

\usepackage[dvipsnames]{xcolor} 

\usepackage[breaklinks,colorlinks,citecolor=blue,urlcolor=blue]{hyperref}
\usepackage{orcidlink}

\newcommand{\gns}{Gaia NS1 }
\newcommand{\gn}{Gaia NS1}

\renewcommand{\arraystretch}{1.2}  
\usepackage{listings}
\usepackage{color}
\definecolor{dkgreen}{rgb}{0,0.6,0}
\definecolor{gray}{rgb}{0.5,0.5,0.5}
\definecolor{mauve}{rgb}{0.58,0,0.82}
\definecolor{golden}{rgb}{0.86,0.65,0.01}
\begin{document}


\title{A $1.9\,M_{\odot}$ neutron star candidate in a 2-year orbit}

\author{\vspace{-1.0cm}Kareem El-Badry\,\orcidlink{0000-0002-6871-1752}$^{1,2}$}
\author{Joshua D. Simon\,\orcidlink{0000-0002-4733-4994}$^{3}$}
\author{Henrique Reggiani\,\orcidlink{0000-0001-6533-6179}$^{4,3}$}
\author{Hans-Walter Rix\,\orcidlink{0000-0003-4996-9069}$^{2}$}
\author{David W. Latham\,\orcidlink{0000-0001-9911-7388}$^{5}$}
\author{Allyson Bieryla\,\orcidlink{0000-0001-6637-5401}$^{5}$}
\author{Lars A. Buchhave\,\orcidlink{0000-0003-1605-5666}$^{6}$}
\author{Sahar Shahaf\,\orcidlink{0000-0001-9298-8068}$^{7}$}
\author{Tsevi Mazeh\,\orcidlink{0000-0002-3569-3391}$^{8}$}
\author{Sukanya Chakrabarti\,\orcidlink{0000-0001-6711-8140}$^{9}$}
\author{Puragra Guhathakurta\,\orcidlink{0000-0001-8867-4234}$^{10}$}
\author{Ilya V. Ilyin\,\orcidlink{0000-0002-0551-046X}$^{11}$}
\author{Thomas~M.~Tauris\,\orcidlink{0000-0002-3865-7265}$^{12}$}

\affiliation{$^1$Department of Astronomy, California Institute of Technology, 1200 E. California Blvd., Pasadena, CA 91125, USA}
\affiliation{$^2$Max-Planck Institute for Astronomy, K\"onigstuhl 17, D-69117 Heidelberg, Germany}
\affiliation{$^3$Observatories of the Carnegie Institution for Science, 813 Santa Barbara St., Pasadena, CA 91101, USA}
\affiliation{$^4$Gemini Observatory/NSF’s NOIRLab, Casilla 603, La Serena, Chile}
\affiliation{$^5$Center for Astrophysics ${\rm \mid}$ Harvard \& Smithsonian,  60 Garden Street, Cambridge, MA 02138, USA}
\affiliation{$^6$DTU Space, National Space Institute, Technical University of Denmark, Elektrovej 328, DK-2800 Kgs. Lyngby, Denmark}
\affiliation{$^7$Department of Particle Physics and Astrophysics, Weizmann Institute of Science, Rehovot 7610001, Israel}
\affiliation{$^8$School of Physics and Astronomy, Tel Aviv University, Tel Aviv, 6997801, Israel}
\affiliation{$^9$Department of Physics and Astronomy, University of Alabama, Huntsville, 301 North Sparkman Drive, Huntsville, USA}
\affiliation{$^{10}$University of California Santa Cruz, UC Observatories, 1156 High Street, Santa Cruz, CA 95064, USA}
\affiliation{$^{11}$Leibniz-Institut für Astrophysik Potsdam (AIP), An der Sternwarte 16, D-14482 Potsdam, Germany}
\affiliation{$^{12}$Department of Materials and Production, Aalborg University, Skjernvej 4A, DK-9220~Aalborg {\O}st, Denmark}
\email{Corresponding author: kelbadry@caltech.edu}

\begin{abstract}
We report discovery and characterization of a main-sequence G star orbiting a dark object with mass $1.90\pm 0.04\,M_{\odot}$. The system was discovered via {\it Gaia} astrometry and has an orbital period of 731 days. We obtained multi-epoch RV follow-up over a period of 639 days, allowing us to refine the {\it Gaia} orbital solution and precisely constrain the masses of both components. The luminous star is a $\gtrsim 12$\,Gyr-old, low-metallicity halo star near the main-sequence turnoff ($T_{\rm eff} \approx 6000$\,K; $\log\left(g/\left[{\rm cm\,s^{-2}}\right]\right)\approx 4.0$; $\rm [Fe/H]\approx-1.25$; $M\approx0.79\,M_{\odot}$) with a highly enhanced lithium abundance. The RV mass function sets a minimum companion mass for an edge-on orbit of $M_2 > 1.67\,M_{\odot}$, well above the Chandrasekhar limit. The {\it Gaia} inclination constraint, $i=68.7\pm 1.4$\,deg, then implies a companion mass of $M_2= 1.90\pm 0.04\,M_{\odot}$. The companion is most likely a massive neutron star: the only viable alternative is two massive white dwarfs in a close binary, but this scenario is disfavored on evolutionary grounds. 
The system's low eccentricity ($e=0.122\pm 0.002$) disfavors dynamical formation channels and implies that the neutron star likely formed with little mass loss ($\lesssim 1\,M_{\odot}$) and with a weak natal kick ($v_{\rm kick}\lesssim 20\,\rm km\,s^{-1}$). Stronger kicks with more mass loss are not fully ruled out but would imply that a larger population of similar systems with higher eccentricities should exist. The current orbit is too small to have accommodated the neutron star progenitor as a red supergiant or super-AGB star. The simplest formation scenario -- isolated binary evolution -- requires the system to have survived unstable mass transfer and common envelope evolution with a donor-to-accretor mass ratio $>10$. The system, which we call \gn, is likely a progenitor of symbiotic X-ray binaries and long-period millisecond pulsars. Its discovery challenges binary evolution models and bodes well for {\it Gaia's} census of compact objects in wide binaries.

\end{abstract}

\maketitle

\section{Introduction}
\label{sec:intro}
Astrometry from the {\it Gaia} mission \citep{GaiaCollaboration2016} has opened a new window on the population of compact stellar remnants lurking in non-interacting binaries. Precise measurements of astrometric ``wobble'' over the course of several years allow {\it Gaia} to detect the presence of binary companions via their gravitational effects, even when they do not emit any light \citep{GaiaCollaboration2023}. Unlike most other binary detection methods, astrometry is most sensitive to companions that are dark and in wide orbits ($P_{\rm orb}\sim 1000$\,d for the currently published data). 

Astrometric orbital solutions from the mission's third data release \citep[DR3;][]{Vallenari2023} have already enabled the discovery of thousands of white dwarfs \citep[WDs;][]{Shahaf2023, Yamaguchi2023, Shahaf2023b} and two black holes \citep[BHs;][]{El-Badry2023, Chakrabarti2023, El-Badry2023_bh2} orbited by luminous stars. The orbits of both the BH and WD systems are unexpectedly wide, falling in a period range that binary evolution models predict to be sparsely populated. BH and WD companions in au-scale orbits appear not to be rare compared to closer binaries. They were  under-represented in samples before {\it Gaia} because discoveries based on photometric variability, radial velocity (RV) shifts, and signatures of accretion all favor short-period systems.

Unlike BHs and WDs, neutron star (NS) companions have not yet been unambiguously identified from the {\it Gaia} data. Several factors make astrometric identification of NSs difficult. The mass distribution of well-characterized NSs is peaked near $1.3\,M_{\odot}$ \citep{Ozel2012, Kiziltan2013, Ozel2016}, below the maximum WD mass  \citep[$\sim 1.38\,M_{\odot}$;][]{Nomoto1987}. This means that a majority of NSs cannot be distinguished unambiguously from high-mass WDs on the basis of their mass alone. Even for NSs with masses above the maximum WD mass, observational uncertainties generally make mass constraints from {\it Gaia} data alone consistent with being below the maximum WD mass at the $(1-3)\sigma$ level \citep[e.g.,][]{Shahaf2023}. Finally, most NSs are thought to form with strong birth kicks due to asymmetric supernova explosions \citep[e.g.,][]{Hobbs2005, Janka2012}. These kicks -- coupled with a sudden drop in binaries' binding energy due to mass loss -- likely disrupt a large majority of would-be NS binaries. It is probably largely for this reason that fewer than 1\% of all known young pulsars are in binaries \citep[e.g.,][]{Lorimer2008}, even though most of the massive stars from which these pulsars form are in binaries and higher-order multiples \citep[e.g.,][]{sana2012}. 

After the release of {\it Gaia} DR3, we initiated an RV follow-up program targeting $\sim 50$ binaries with astrometric solutions suggesting the companion might be a NS. RV follow-up of these candidates is ongoing and described in a companion paper, El-Badry et al. (2024, in prep). In this paper, we present a detailed analysis of one of the most interesting candidates from that sample, which we refer to as \gns ({\it Gaia} DR3 source ID 6328149636482597888). Besides \gn, our follow-sample includes 20 other high-quality NS candidates.

Compared to other candidates we have followed-up, \gns is unique in three respects: (a) it is the only system for which RVs alone set a minimum mass for the dark companion that is unambiguously above the Chandrasekhar mass, (b) it has the highest inferred dark companion mass of all NS candidates we have followed up, and (c) it has the lowest eccentricity among the NS candidates. These features  make it unlikely that the unseen companion is anything other than a NS, and they make the system's properties particularly difficult to explain with binary evolution models. 

The remainder of this paper is organized as follows. Section~\ref{sec:disc} describes how the system was identified as a compact object candidate, while Section~\ref{sec:properties} summarizes the source's basic observational properties. Section~\ref{sec:spectra} describes our spectroscopic observations and measurement of RVs, while Section~\ref{sec:rvfit} describes fitting of the RVs and {\it Gaia} data to constrain the companion mass. In Section~\ref{sec:spec_anal}, we measure atmospheric parameters and abundances of the luminous star. Section~\ref{sec:galpy} examines the system's Galactic orbit. We discuss the nature of the companion and the system's possible formation histories in Section~\ref{sec:discussion_kicks} and conclude in Section~\ref{sec:conclusion}.

\section{Discovery}
\label{sec:disc}
\gns was identified as a compact object candidate soon after {\it Gaia} DR3 by several authors. \citet{Andrews2022} assumed a luminous star mass of $M_\star = 1.21\pm 0.2\,M_{\odot}$ and calculated a dark companion mass of $M_2=2.71_{-0.36}^{+1.50}\,M_{\odot}$ (2$\sigma$ uncertainties) based on the astrometric orbit. They classified the unseen companion as a BH. \citet{Shahaf2023} assumed $M_\star=1.09\,M_{\odot}$  and inferred $M_2 = 2.45\pm 0.2\,M_{\odot}$, also classifying the companion as a BH. The luminous star mass estimate adopted by \citet{Shahaf2023} was from  the \texttt{gaiadr3.binary\_masses} table \citep{GaiaCollaboration2023}, while the estimate adopted by \citet{Andrews2022} was from the {\it Gaia} FLAME estimator \citep{Creevey2023}. Both values are based on comparison of the source's color and absolute magnitude to isochrones, with a metallicity prior that favors solar metallicity. 

\citet{El-Badry2023} reported a spectroscopic metallicity of $[\rm Fe/H]\approx -1.5$ for the luminous star and a revised mass estimate of only $M_{\star}=0.78\pm 0.05$ based on a metallicity-informed fit of the spectral energy distribution. This significantly lower mass estimate reflects the facts that (a) lower-metallicity stars are brighter at fixed mass, and (b) the star's CMD position implies that it is near the main-sequence turnoff, having already expanded significantly since the zero age main sequence. Given this mass estimate for the luminous star, they estimated $M_{2}=2.25_{-0.26}^{+0.61}M_{\odot}$ based on the {\it Gaia} astrometric solution. 

\citet{El-Badry2023_bh2} subsequently reported that early radial velocity (RV) measurements were roughly consistent with the {\it Gaia} astrometric solution but cautioned that RV coverage over a majority of the orbit would be required to validate or refute the astrometric solution. We have now obtained the necessary RVs.

\section{Observational properties }
\label{sec:properties}
\gns is a bright ($G=13.3$) source in the Libra constellation ($\alpha=$\,14:32:20.7; $\delta=-$10:21:59)
at high Galactic latitude ($l=339.5$, $b=45.2$). The field is sparse, with no other {\it Gaia} sources within 12 arcsec. The {\it Gaia} DR3 astrometric solution has a proper motion of 92\,mas\,yr$^{-1}$ and a parallax of $1.23\pm 0.04$ mas. This implies a tangential velocity of $\approx 350\,\rm km\,s^{-1}$, characteristic of a halo star. On the color-magnitude diagram (Figure~\ref{fig:inc_fig}), the source falls on top of the solar neighborhood main sequence. It has a de-reddened absolute magnitude $M_{G,0}\approx 3.5$, about twice as bright as the Sun. This led \citet{Andrews2022} and \citet{Shahaf2023} to assume masses of $M_\star \approx 1.1-1.2\,M_{\odot}$. As we show below, the object is a low-metallicity turn-off star with a somewhat lower mass.

\subsection{Distance}
\label{sec:dist}
The parallax constraint from \gn's DR3 astrometric binary solution is $\varpi=1.23\pm 0.04$\,mas, corresponding to a distance of $d\approx 812\pm 26$\,pc. However, the parallax is covariant with other parameters of the astrometric model, and our joint fitting of RVs and astrometric constraints in Section~\ref{sec:rvfit} leads to revised constraints on those parameters, thus also updating the parallax constraint. The parallax constraint from our joint fit is $\varpi = 1.36\pm 0.03$ mas, corresponding to a closer distance of $d\approx 735$\,pc. 

We adopt the parallax from the joint fit and associated closer distance of 735 pc in most of our fiducial modeling. We also explore how a larger distance of $\approx 812$ pc would change our results, finding that it would change the best-fit masses of both components by less than $2\%$. We discuss the modest tension between the {\it Gaia}-only and {\it Gaia}+RV parallaxes in Section~\ref{sec:consistency}.

\subsection{Spectral energy distribution}
\label{sec:sed}
 
We constructed the source's broadband spectral energy distribution (SED) by combining UV photometry from {\it GALEX} \citep{Martin2005}, synthetic SDSS $ugriz$  photometry constructed from {\it Gaia} BP/RP spectra \citep{GaiaCollaboration2022}, 2MASS $JHK$ photometry \citep[][]{Skrutskie_2006}, and WISE $W1\,W2\,W3$ photometry \citep[][]{Wright_2010}. We set an uncertainty floor of 0.03 mag in all bands to account for photometric calibration issues and imperfect models. We then fit the SED with single-star models, with a prior on the metallicity and $\alpha$-abundance from spectroscopy ($[\rm Fe/H]=-1.23\pm 0.08$ and $\rm [\alpha/Fe]=0.20$, see Section~\ref{sec:spec_anal}). 

We predict bandpass-integrated magnitudes using empirically calibrated model spectral from the BaSeL library \citep[v2.2;][]{Lejeune1997, Lejeune1998}. We assume a \citet{Cardelli_1989} extinction law with $R_V =3.1$ and adopt a prior on the reddening $E(B-V) = 0.08\pm 0.02$ based on the \citet{Green2019} 3D dust map. We use \texttt{pystellibs}\footnote{\href{https://mfouesneau.github.io/pystellibs/}{https://mfouesneau.github.io/pystellibs/}} to interpolate between model SEDs, and \texttt{pyphot}\footnote{\href{https://mfouesneau.github.io/pyphot/}{https://mfouesneau.github.io/pyphot/}}  to calculate synthetic photometry.

We fit the SED using \texttt{emcee} \citep{emcee2013} to sample from the posterior, with the initial mass, metallicity, age, parallax, and reddening sampled as free parameters. We use the \texttt{MINESweeper} framework \citep{Cargile2020} to interpolate on MIST isochrones \citep{Choi2016} to predict the radius, effective temperature, and surface metallicity from the initial mass, age, and metallicity. These calculations account for atomic diffusion, which causes the present-day surface metallicity to be somewhat lower than the star's average metallicity \citep[e.g.,][]{Dotter2017}. We set a maximum age of 13.5\,Gyr.

\subsection{Luminous star mass constraints}
\label{sec:lum_mass}

\begin{figure*}
    \centering
    \includegraphics[width=\textwidth]{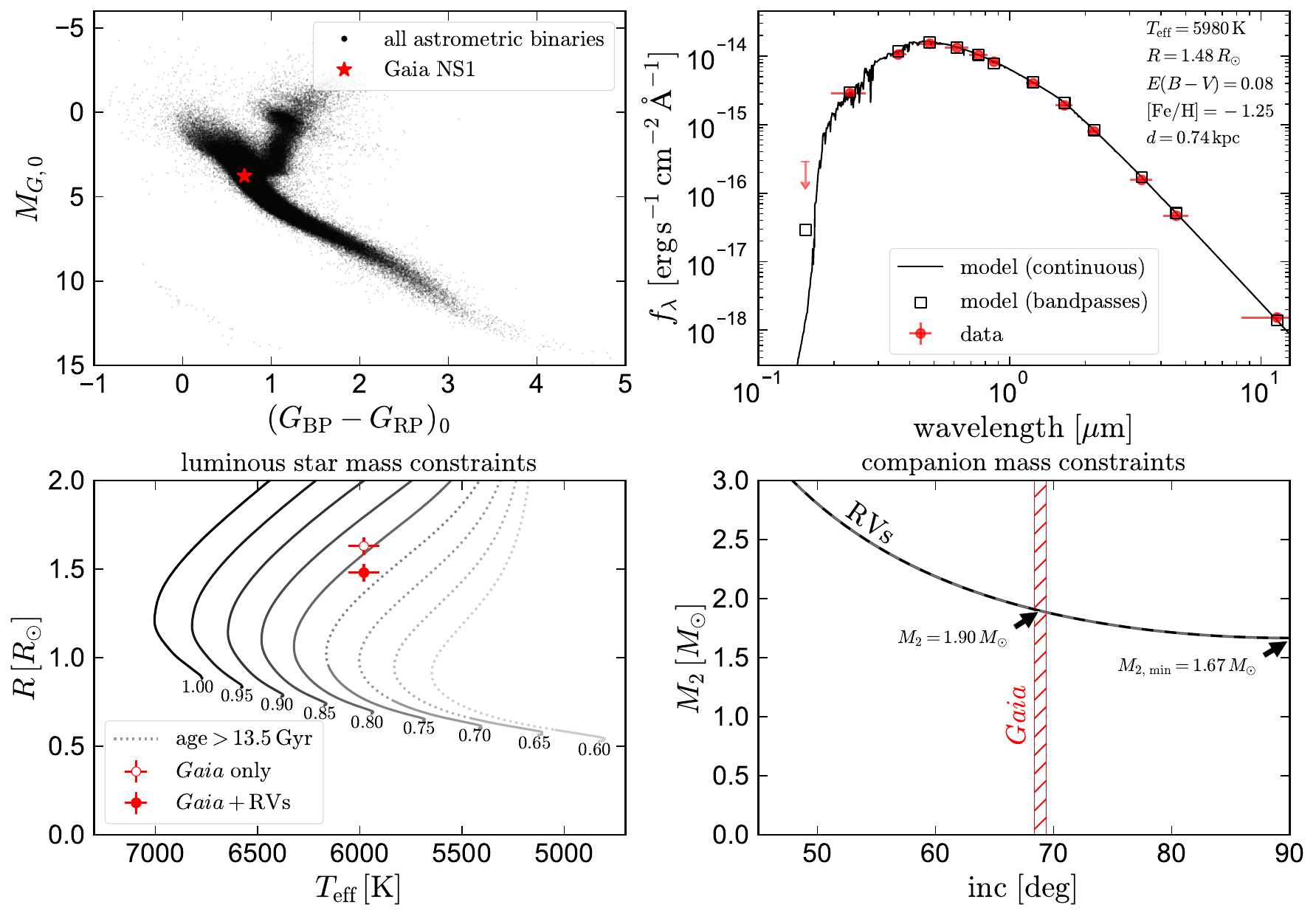}
    \caption{Observational properties and mass constraints of \gn. Upper left: extinction-corrected color--magnitude diagram. Gaia NS1 falls on the main sequence and is slightly bluer and brighter than the Sun. Upper right: broadband spectral-energy distribution. Red points show observed photometry; open black squares show the integrated fluxes predicted for the best-fit model spectrum (black line). 
    Lower left: comparison of the measured temperature and radius to MIST single-star evolutionary models. Solid point shows the radius inferred when adopting the parallax from our joint {\it Gaia} + RV fit; hollow point shows the parallax from {\it Gaia} alone (see Section~\ref{sec:dist}). The observed temperature, radius, and metallicity imply a luminous star mass of $0.79\pm 0.01\,M_{\odot}$.  Lower right: dynamical constraints on the companion mass. The RVs and luminous star mass constrain the companion mass and inclination to the gray shaded region, with a minimum companion mass of $1.67\,M_{\odot}$ for an edge-on orbit. The inclination constraint from {\it Gaia} astrometry then implies $M_2 = 1.90\,M_{\odot}$.}
    \label{fig:inc_fig}
\end{figure*}

The results of SED fitting are shown in Figure~\ref{fig:inc_fig} and reported in Table~\ref{tab:system}. We find a luminous star mass of $M_\star =0.79\pm 0.01\,M_{\odot}$ and an age of $12.5\pm 0.5$\,Gyr. The age posterior samples extend to the upper limit of the prior, with a 2$\sigma$ lower-limit of 11.5\,Gyr. Given the star's low metallicity and halo-like orbit (Section~\ref{sec:galpy}), an age of 11+ Gyr is expected \citep[e.g.,][]{Xiang2022}.

The inferred parameters place the star at the end of its main-sequence evolution and imply that it will soon become a red giant, reaching the tip of the giant branch within $\lesssim$1 Gyr. The current radius, $R\approx 1.48\,R_{\odot}$, is already a factor of $\sim 2.1$ larger than the value predicted for the same mass at the zero-age main sequence. The MIST evolutionary models predict that atomic diffusion has significantly reduced the surface metallicity: while they predict a current surface iron abundance $[\rm Fe/H] \approx -1.23$, the corresponding initial metallicity is $[\rm Fe/H]\approx -1.08$. Accounting for the star's enhanced abundance of $\alpha$ elements ($[\rm \alpha/Fe]\approx +0.20$; Section~\ref{sec:spec_anal}), this implies an initial bulk metallicity of $[\rm M/H] \approx -0.95$. 

We show the luminous star's inferred temperature and radius in the lower left panel of Figure~\ref{fig:inc_fig} with a solid red point. The hollow red point shows the parameters inferred when we adopt the {\it Gaia}-only parallax: in that case, the $\approx 10\%$ larger distance results in a 10\% larger inferred radius. We compare the measured parameters to MIST evolutionary tracks with initial $[\rm M/H] = -0.95$. 

Because the models predict rapid evolution near the main-sequence turnoff, the inferred mass and age when we adopt the {\it Gaia}-only parallax are only slightly different: $M_\star = 0.81 \pm 0.015\,M_{\odot}$ and $\tau = 12.0\pm 0.7$\,Gyr, still corresponding to a star near the main-sequence turnoff.  The tracks in Figure~\ref{fig:inc_fig} also demonstrate that -- even if our adopted distance were grossly in error -- there is no plausible scenario under single-star evolution in which the mass of the luminous star is below $0.75\,M_{\odot}$: models with $M< 0.75\,M_{\odot}$ do not reach $T_{\rm eff} \approx 6000$\,K at any age below 13.5 Gyr. 

There is good agreement between the inferred mass, radius, surface gravity, and distance. We discuss whether binary evolution could have produced a lower-mass luminous star with the observed parameters in Section~\ref{sec:lower_mass}, concluding that such a scenario is very unlikely.

\section{Spectroscopic observations}
\label{sec:spectra}

\begin{table}
\begin{tabular}{lll}
HJD UTC & RV ($\rm km\,s^{-1}$) & Instrument  \\
\hline
2459767.5238 & $130 \pm 3$ & MagE \\
2459795.5592 & $136 \pm 3$ & MagE \\
2459813.4918 & $139.74 \pm 0.10$ & FEROS \\
2459831.5038 & $143.61 \pm 0.41$ & FEROS \\
2459832.4866 & $143.60 \pm 0.25$ & FEROS \\
2459929.0525 & $154.65 \pm 0.50$ & PEPSI \\
2459977.8497 & $154.02 \pm 0.05$ & MIKE \\
2459981.0076 & $153.69 \pm 0.08$ & TRES \\
2459990.8528 & $153.22 \pm 0.15$ & FEROS \\
2459992.8178 & $153.16 \pm 0.15$ & FEROS \\
2460012.8160 & $151.78 \pm 0.12$ & FEROS \\
2460026.7002 & $150.29 \pm 0.15$ & FEROS \\
2460036.8989 & $149.37 \pm 0.07$ & TRES \\
2460037.6527 & $149.41 \pm 0.04$ & MIKE \\
2460038.7825 & $149.18 \pm 0.02$ & MIKE \\
2460039.7097 & $149.14 \pm 0.09$ & FEROS \\
2460044.7941 & $148.47 \pm 0.08$ & APF \\
2460044.8121 & $148.70 \pm 0.09$ & APF \\
2460046.9207 & $148.36 \pm 0.07$ & TRES \\
2460050.6922 & $147.89 \pm 0.10$ & FEROS \\
2460054.8601 & $147.23 \pm 0.11$ & APF \\
2460054.8803 & $147.34 \pm 0.13$ & APF \\
2460057.7359 & $147.01 \pm 0.05$ & MIKE \\
2460059.8472 & $146.89 \pm 0.10$ & TRES \\
2460061.8931 & $146.48 \pm 0.09$ & APF \\
2460061.9128 & $146.64 \pm 0.08$ & APF \\
2460072.7468 & $145.13 \pm 0.20$ & FEROS \\
2460072.8152 & $145.07 \pm 0.23$ & APF \\
2460072.8354 & $145.40 \pm 0.13$ & APF \\
2460076.6110 & $144.73 \pm 0.04$ & MIKE \\
2460078.7900 & $144.35 \pm 0.16$ & TRES \\
2460085.6962 & $143.28 \pm 0.22$ & FEROS \\
2460086.8885 & $143.20 \pm 0.16$ & APF \\
2460086.9084 & $143.62 \pm 0.15$ & APF \\
2460094.8017 & $142.60 \pm 0.08$ & TRES \\
2460095.6516 & $142.37 \pm 0.03$ & MIKE \\
2460098.6573 & $141.94 \pm 0.25$ & FEROS \\
2460106.8061 & $140.97 \pm 0.17$ & APF \\
2460109.7473 & $140.43 \pm 0.06$ & TRES \\
2460110.6418 & $140.22 \pm 0.07$ & FEROS \\
2460122.7615 & $138.51 \pm 0.09$ & APF \\
2460122.7810 & $138.64 \pm 0.11$ & APF \\
2460132.5799 & $136.99 \pm 0.08$ & MIKE \\
2460133.6779 & $136.95 \pm 0.07$ & TRES \\
2460139.5731 & $136.22 \pm 0.13$ & FEROS \\
2460168.5282 & $131.85 \pm 0.08$ & MIKE \\
2460185.4878 & $128.99 \pm 0.21$ & FEROS \\
2460186.4889 & $128.99 \pm 0.15$ & FEROS \\
2460303.0427 & $114.39 \pm 0.32$ & APF \\
2460306.0444 & $113.88 \pm 0.12$ & TRES \\
2460327.0182 & $112.34 \pm 0.07$ & TRES \\
2460338.8541 & $111.97 \pm 0.10$ & FEROS \\
2460340.8396 & $111.67 \pm 0.08$ & FEROS \\
2460345.8491 & $111.44 \pm 0.10$ & MIKE \\
2460352.9597 & $111.21 \pm 0.08$ & TRES \\
2460374.9193 & $111.29 \pm 0.09$ & TRES \\
2460386.9141 & $111.63 \pm 0.09$ & TRES \\
2460400.8157 & $112.42 \pm 0.06$ & FEROS \\
2460406.6343 & $112.81 \pm 0.12$ & MIKE \\
\hline \hline
\multicolumn{2}{l}{\bf{RV zeropoint offsets relative to FEROS}}   \\ 
Instrument & offset ($\rm km\,s^{-1}$)\\
\hline 
TRES & $0.15 \pm 0.01$ \\ 
MIKE & $0.36 \pm 0.01$\\ 
APF & $-0.86 \pm 0.02$ \\ 
\hline 
\end{tabular}
\caption{RVs. The best-fit instrumental offsets (lower block) have been subtracted so that all RVs are reported on the same scale.}
\label{tab:rvs}
\end{table}

\subsection{Data}
The promise of \gns as a NS or BH binary prompted follow-up at several telescopes and by several groups, which are synthesized into a coherent analysis here.

\subsubsection{FEROS}
\label{sec:feros}
We observed \gns 19 times using the Fiberfed Extended Range Optical Spectrograph \citep[FEROS;][]{Kaufer1999} on the 2.2m ESO/MPG telescope at La Silla Observatory (programs P109.A-9001, P110.A-9014, P111.A-9003, and P112.A-6010). The spectra have resolution $R\approx 50,000$ and cover 360--920 nm. Most of our observations used 2400\,s exposures. We reduced the data using the CERES pipeline \citep{Brahm2017}, which performs bias-subtraction, flat fielding, wavelength calibration, and optimal extraction. The pipeline measures and corrects for small shifts in the wavelength solution during the course a night via simultaneous observations of a ThAr lamp with a second fiber. For RV measurements (Section~\ref{sec:RVs}), we used 15 orders with wavelengths between 4500 and 6700\,\AA.

\subsubsection{TRES}
\label{sec:tres}
We observed \gns 13 times  using the Tillinghast Reflector Echelle Spectrograph \citep[TRES;][]{Furesz2008} mounted on the 1.5 m Tillinghast Reflector telescope at the Fred Lawrence Whipple Observatory (FLWO) on Mount Hopkins, Arizona. TRES is a fibrefed echelle spectrograph with a wavelength range of 390--910 nm and a resolving power of $R\sim 44,000$. The spectra were extracted as described in \citet{Buchhave2010}. We measured RVs using 31 orders covering the wavelength range of 420--670\,nm.

\subsubsection{APF}
\label{sec:apf}
We observed \gns 14 times using the Levy spectrometer on the 2.4 m Automated Planet Finder Telescope \citep[APF;][]{Radovan2010, Vogt2014} at Lick Observatory. We used a 2$\arcsec$ slit, yielding spectra with resolution $R\approx 80,000$ and coverage over 373–1020\,nm. Our observations did not use the iodine cell. We reduced the spectra using the California Planet Search pipeline \citep{Howard2010, Fulton2015} and measured RVs using 35 orders orders covering the wavelength range of 440--670\,nm.

\subsubsection{MIKE}
\label{sec:MIKE}

We observed \gns 10 times with the Magellan Inamori Kyocera Echelle (MIKE) spectrograph on the Magellan Clay telescope at Las Campanas Observatory \citep{Bernstein2003SPIE}. We used the 0.5$\arcsec$ slit for 3 observations, and the 0.7$\arcsec$ slit for the other 6. Exposure times ranged from 300 to 2160s, yielding spectral resolution $R \sim 40,000$ on the blue side and $R \sim 55,000$ on the red side. The total wavelength coverage was 333--968\,nm. 
We reduced the spectra with the MIKE pipeline within \texttt{CarPy} \citep{Kelson2000ApJ, Kelson2003PASP}.  We measured RVs using 18 orders on the red side covering the wavelength range of 500--680\,nm.

\subsubsection{MagE}
\label{sec:MagE}
We observed \gns twice with the Magellan Echellette spectrograph \citep[MagE;][]{Marshall2008} on the 6.5m Magellan Clay telescope at Las Campanas Observatory. Both observations were carried out with the 0.7 arcsec slit and a 300\,s exposure. This yielded  spectral resolution $R\approx 5,400$ and wavelength coverage of 350--1100\,nm. We reduced the spectra using \texttt{PypeIt} \citep[][]{Prochaska_2020}.

\subsubsection{PEPSI}
\label{sec:PEPSI}
We obtained one spectrum  using the Potsdam Echelle Polarimetric and Spectroscopic Instrument (PEPSI; \citealt{Strassmeier2015}) spectrograph on the Large Binocular Telescope in binocular mode. We used the 300\,$\mu$m fiber and the CD2 and CD5 cross-dispersers on the blue and red side, respectively, with an exposure time of 600\,s. The spectrum was reduced  as described in \citet{Strassmeier2018}; it covers the wavelength range of 422--479\,nm and 624-743\,nm with spectral resolution $R\approx 50,000$.

\subsection{Radial velocity measurements}
\label{sec:RVs}
We measured RVs from all the spectra by cross-correlating the individual orders with a Kurucz template spectrum from the BOSZ library \citep{Bohlin2017}. Motivated by the spectroscopic analysis (Section~\ref{sec:spec_anal}), we used a template with $T_{\rm eff} = 6000$\,K, $\log g = 4$, and $\rm [Fe/H] = -1.25$. For most instruments, we calculated uncertainties on the RVs from the standard deviation of the RVs inferred from individual orders divided by the square root of the number of orders used in the analysis. For the MagE spectra, where  the uncertainty associated with the telluric RV correction is expected to be substantial due to lower resolution, we adopted conservative uncertainties of $3\,\rm km\,s^{-1}$. 

To correct for shifts in the wavelength solution due to instrumental flexure, slit centering errors, and other systematics, we cross-correlated a telluric model spectrum synthesized with \texttt{TelFit} \citep{Gullikson2014} with the telluric ``A'' and ``B'' bands due to molecular oxygen in the observed spectra \citep{Griffin1973}. We applied the thus-inferred corrections to the RVs measured from the APF, MIKE, MagE, and PEPSI spectra, where they reached a maximum amplitude of a few $\rm km\,s^{-1}$. We did not apply them to the FEROS and TRES spectra, where the best-fit shifts had typical amplitudes of a few tens of $\rm m\,s^{-1}$, comparable to their uncertainties. We applied heliocentric corrections to all RVs  after the telluric corrections.

All our measured RVs are listed in Table~\ref{tab:rvs}. The median uncertainty is 0.10\,$\rm km\,s^{-1}$. This is slightly larger than typical for our follow-up, due mainly to the star's low metallicity, but still $400+$ times smaller than the star's RV variability amplitude, allowing the orbit to be tightly constrained.

\begin{table}
\centering
\caption{Physical parameters and 1$\sigma$ uncertainties on the properties of the luminous star. We separately list basic observables (1st block), constraints from SED fitting (2nd block; Section~\ref{sec:lum_mass}), and constraints from the hybrid spectroscopic + isochrone fit (Section~\ref{sec:spec_anal}; 3rd block).}
\begin{tabular}{lll}
\hline\hline
\multicolumn{3}{l}{\bf{Properties of the unresolved source}}   \\ 
Right ascension & $\alpha$\,[deg] & 218.08620 \\
Declination & $\delta$\,[deg] & -10.36635 \\
Apparent magnitude & $G$\,[mag] & 13.34 \\
Extinction & $E(B-V)$\,[mag] & $ 0.08 \pm 0.02 $ \\

\hline
\multicolumn{3}{l}{\bf{Parameters of the luminous star (SED fit)}}  \\ 
Effective temperature & $T_{\rm eff}$\,[K] & $5980 \pm 55 $ \\
Radius & $R_\star\,[R_{\odot}]$ & $1.48 \pm 0.05$  \\ 
Bolometric luminosity & $L_\star\,[L_{\odot}]$ & $2.51\pm 0.15$ \\ 
Mass &  $M_\star\,[M_{\odot}]$ & $0.79 \pm 0.01$ \\
Age &  $\tau\,\rm [Gyr]$ & $12.5 \pm 0.5$ \\ 
\hline
\multicolumn{3}{l}{\bf{Parameters of the luminous star (spectroscopic fit)}}  \\ 
Effective Temperature & $T_{\rm eff}$ [K] & $6077^{+15}_{-16}$ \\
Surface gravity   & $\log(g/(\rm cm\,s^{-2}))$ & $4.02 \pm 0.01 $ \\
Metallicity & [Fe/H] & $-1.23 \pm 0.08 $ \\
Microturbulent velocity & $\xi$ [km s$^{-1}$] & $0.8 \pm 0.2 $ \\
Mass & $M_\star\,[M_{\odot}]$ & $0.80 \pm 0.02 $ \\
Age & $\tau$ [Gyr] & $13.2^{+0.2}_{-0.4}$ \\
Projected rotation velocity & $v\sin i$\,[km\,s$^{-1}$] &  $< 3.0$ \\
Abundance pattern & [X/Fe]&  Table~\ref{tab:abund} \\

\hline
\end{tabular}
\begin{flushleft}

\label{tab:star}
\end{flushleft}
\end{table}

\begin{table}
\centering
\caption{Physical parameters and 1$\sigma$ uncertainties of \gn's orbit. We compare constraints on the orbit based on both {\it Gaia} and our RVs (1st block), the {\it Gaia} solution alone (2nd block), and our RVs alone (3rd and 4th block; in the 4th block, the period is fixed to the value inferred from the joint fit). Constraints with the {\it Gaia} uncertainties inflated by a factor of 3 are listed in the 5th block. We consider these the most robust constraints. }
\begin{tabular}{lll}
\hline\hline

\hline
\multicolumn{3}{l}{\bf{Parameters of the orbit ({\it Gaia} + RVs)}}  \\ 
Orbital period & $P_{\rm orb}$\,[days] & $ 730.9 \pm 0.5 $ \\
Semi-major axis & $a$\,[au] & $ 2.21 \pm 0.006$ \\
Photocenter semi-major axis & $a_0$\,[mas] & $ 2.130 \pm 0.015 $ \\
Eccentricity & $e$ & $ 0.123 \pm 0.002 $ \\
Inclination  & $i$\,[deg] & $ 68.8 \pm 0.5 $ \\
Periastron time & $T_p$\,[JD-2457389] & $187.0 \pm 1.9 $ \\
Ascending node angle & $\Omega$\,[deg] & $82.6 \pm 0.8 $ \\
Argument of periastron & $\omega$\,[deg] & $259.7 \pm 0.5 $ \\
Neutron star mass & $M_2$\,[$M_{\odot}$] & $1.902 \pm 0.015$ \\
Center-of-mass RV & $\gamma$\,[$\rm km\,s^{-1}$] & $133.46 \pm 0.03$ \\
RV semi-amplitude & $K_\star$\,[$\rm km\,s^{-1}$] & $21.83 \pm 0.03$ \\
RV mass function & $f\left(M_{2}\right)_{{\rm RVs}}$\,[$\rm M_{\odot}$] & $0.770 \pm 0.003$ \\
Parallax & $\varpi$\,[mas] & $ 1.36 \pm 0.01 $ \\
Proper motion in RA & $\mu_{\alpha}^{*}$\,[$\rm mas\,yr^{-1}$] & $ -0.01 \pm 0.02 $ \\
Proper motion in Dec & $\mu_{\delta}$\,[$\rm mas\,yr^{-1}$] & $ -92.36 \pm 0.03 $ \\
Tangential velocity & $v_{\perp}\,\left[{\rm km\,s^{-1}}\right]$ & $ 320.8 \pm 2.5$ \\

\hline
\multicolumn{3}{l}{\bf{Parameters of the orbit ({\it Gaia} only)}}  \\ 
Orbital period & $P_{\rm orb}$\,[days] & $ 736 \pm 12 $ \\
Semi-major axis & $a$\,[au] & $ 2.32 \pm 0.06 $ \\
Photocenter semi-major axis & $a_0$\,[mas] & $2.11 \pm 0.07 $ \\
Eccentricity & $e$ & $0.135 \pm 0.035 $ \\
Inclination  & $i$\,[deg] & $70.1 \pm 0.9 $ \\
Periastron time & $T_p$\,[JD-2457389] & $228 \pm 38 $ \\
Ascending node angle & $\Omega$\,[deg] & $83.1 \pm 0.8 $ \\
Argument of periastron & $\omega$\,[deg] & $279 \pm 14 $ \\
Neutron star mass & $M_2$\,[$M_{\odot}$] & $2.27 \pm 0.21 $ \\
RV semi-amplitude & $K_\star$\,[$\rm km\,s^{-1}$] & $24.0 \pm 1.2$ \\
RV mass function & $f\left(M_{2}\right)_{{\rm RVs}}$\,[$\rm M_{\odot}$] & $1.03 \pm 0.15$ \\
Parallax & $\varpi$\,[mas] & $ 1.23 \pm 0.04 $ \\
Proper motion in RA & $\mu_{\alpha}^{*}$\,[$\rm mas\,yr^{-1}$] & $ -0.02 \pm 0.02 $ \\
Proper motion in Dec & $\mu_{\delta}$\,[$\rm mas\,yr^{-1}$] & $ -92.35 \pm 0.06 $ \\
Tangential velocity & $v_{\perp}\,\left[{\rm km\,s^{-1}}\right]$ & $ 355.4 \pm 11.6$ \\ 

\hline
\multicolumn{3}{l}{\bf{Parameters of the orbit (RVs only)}} \\
Orbital period & $P_{\rm orb}$\,[days] & $730.7 \pm 2.2 $ \\
Eccentricity & $e$ & $0.122\pm 0.004$ \\
RV semi-amplitude & $K_\star$\,[$\rm km\,s^{-1}$] & $21.82 \pm 0.04$ \\
RV mass function & $f\left(M_{2}\right)_{{\rm RVs}}$\,[$\rm M_{\odot}$] & $0.769 \pm 0.003$ \\
\hline
\multicolumn{3}{l}{\bf{Joint fit ({\it Gaia} + RVs; $3\times$ inflated {\it Gaia} uncertainties)}}  \\
Orbital period & $P_{\rm orb}$\,[days] & $ 730.8 \pm 1.2 $ \\
Semi-major axis & $a$\,[au] & $ 2.21 \pm 0.01 $ \\
Photocenter semi-major axis & $a_0$\,[mas] & $ 2.13 \pm 0.04 $ \\
Eccentricity & $e$ & $ 0.122 \pm 0.002 $ \\
Inclination  & $i$\,[deg] & $ 68.7 \pm 1.4 $ \\
Periastron time & $T_p$\,[JD-2457389] & $186 \pm 4 $ \\
Ascending node angle & $\Omega$\,[deg] & $82.7 \pm 2.2 $ \\
Argument of periastron & $\omega$\,[deg] & $259.2 \pm 0.6 $ \\
Neutron star mass & $M_2$\,[$M_{\odot}$] & $1.90 \pm 0.04$ \\
Center-of-mass RV & $\gamma$\,[$\rm km\,s^{-1}$] & $133.47 \pm 0.04$ \\
RV semi-amplitude & $K_\star$\,[$\rm km\,s^{-1}$] & $21.82 \pm 0.03$ \\
RV mass function & $f\left(M_{2}\right)_{{\rm RVs}}$\,[$\rm M_{\odot}$] & $0.769 \pm 0.003$ \\
Parallax & $\varpi$\,[mas] & $ 1.36 \pm 0.03 $ \\
Proper motion in RA & $\mu_{\alpha}^{*}$\,[$\rm mas\,yr^{-1}$] & $ -0.01 \pm 0.07 $ \\
Proper motion in Dec & $\mu_{\delta}$\,[$\rm mas\,yr^{-1}$] & $ -92.36 \pm 0.07 $ \\
Tangential velocity & $v_{\perp}\,\left[{\rm km\,s^{-1}}\right]$ & $ 320 \pm 7$ \\ \hline 
\end{tabular}
\begin{flushleft}

\label{tab:system}
\end{flushleft}
\end{table}

\section{Orbit and companion mass}
\label{sec:rvfit}

To explore how the {\it Gaia} astrometry and our follow-up RVs separately constrain the companion mass, we present orbit fits that depend only on RVs (Section~\ref{sec:pure_rv_fixed_period}), on both astrometry and RVs (Section~\ref{sec:joint_fit}), and on astrometry alone (Section~\ref{sec:only_astrometry}). This also allows us to assess the consistency between the astrometry and RVs and to consider the effects of underestimated astrometric uncertainties (Section~\ref{sec:consistency}).

\subsection{Pure RV fit}
\label{sec:pure_rvs}
Neglecting the {\it Gaia} data entirely, we first fit the observed RVs with a Keplerian model. In addition to the standard 7 Keplerian parameters, we fit instrumental offsets for TRES, APF, and MIKE relative to FEROS. We do not fit offsets for MagE and PEPSI, where the small number of RVs would make them poorly constrained. We use \texttt{emcee} \citep{emcee2013} to sample from the posterior, drawing 5000 samples with 64 walkers after a burn-in period of 5000 steps. We initialize the fit at the maximum-likelihood solution implied by the {\it Gaia} constraints assuming a dark companion and use broad, flat priors on all parameters. The RVs are densely sampled and cover most of a full orbit, so the posterior is unimodal. The results are reported in Table~\ref{tab:system} under ``RVs only''. The measured period, $P_{\rm orb}=730.7\pm 2.2$\,d, and eccentricity, $e=0.122\pm 0.004$, are fully consistent with the {\it Gaia} solution. The RV mass function is constrained to $f(M_2)_{\rm RVs} = \frac{P_{{\rm orb}}K_{\star}^{3}}{2\pi G}\left(1-e^{2}\right)^{3/2} = 0.769\pm 0.003\,M_{\odot}$. Constraints on instrumental offsets are reported at the bottom of Table~\ref{tab:rvs}.


\begin{figure*}
    \centering
    \includegraphics[width=\textwidth]{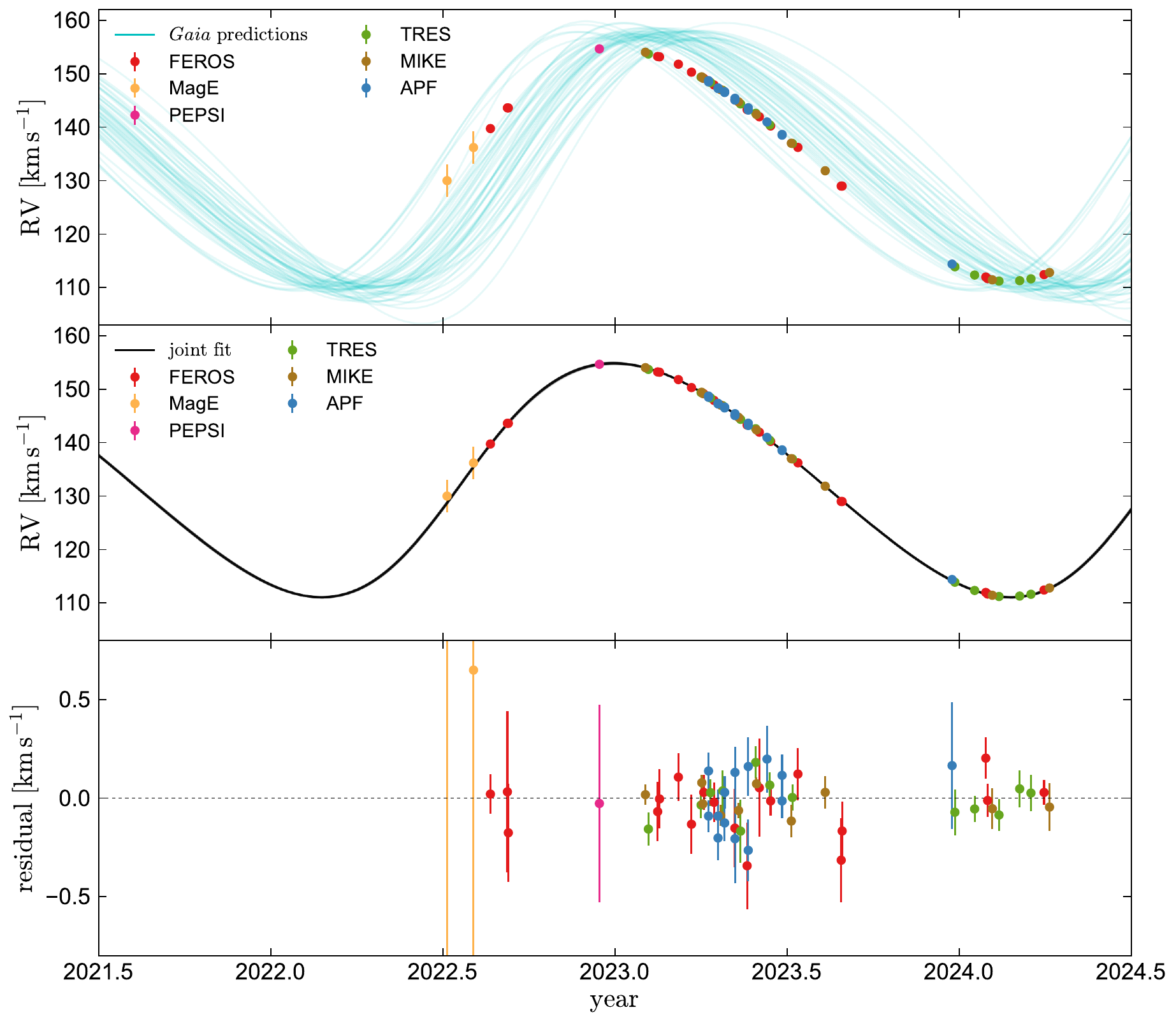}
    \caption{Observed and predicted RVs. In the top panel, cyan lines show predictions of the {\it Gaia} astrometric solution. The center-of-mass RV is set to the value inferred from the joint-fit, but the predictions are otherwise uninformed by the observed RVs. The observed RVs broadly match the behavior predicted by the astrometric solution. In the middle panel, black lines show predictions of a joint fit of the RVs and the astrometric constraints. The combination of RVs and astrometry leads to tight constraints on the orbit. Bottom panel shows residuals of the data compared to the best-fit joint model. These are generally consistent with 0, indicating that the RVs are well-described by a Keplerian orbit.}
    \label{fig:rvfig}
\end{figure*}

\subsection{Joint fit of RVs and astrometry}
\label{sec:joint_fit}

We next jointly fit the RVs and {\it Gaia} astrometry assuming a Keplerian 2-body orbit. The likelihood function is a product of two terms: one comparing the measured and predicted RVs, and another comparing the predicted astrometric orbital parameters to those measured by {\it Gaia}.  The fit has 14 free parameters: orbital period, $P_{\rm orb}$, eccentricity, $e$, luminous star mass,  $M_{\star}$, companion mass, $M_2$, inclination, $i$, ascending node longitude, $\Omega$, argument of periastron, $\omega$, periastron time, $T_{p}$, center-of-mass RV, $\gamma$, parallax, $\varpi$, right ascension, $\alpha$, declination, $\delta$, and proper motions, $\mu_{\alpha}^*$ and $\mu_{\delta}$. We adopt the best-fit instrumental offsets inferred in the RV-only fit rather than re-fitting them. 

Our approach closely follows \citet{El-Badry2023_bh2} and \citet{El-Badry2023}. For each call to the likelihood function, we predict both the RVs of the luminous star at the times of our RV measurements, and the vector of 12 {\it Gaia}-constrained parameters. The latter contains both some free parameters of the fit ($P_{\rm orb}$, $e$, $T_{p}$, $\alpha$, $\delta$, $\varpi$, $\mu_{\alpha}^*$, and $\mu_{\delta}^*$), and some quantities that are transformations of our free parameters (the Thiele-Innes coefficients, $A$, $B$, $F$, and $G$). The astrometric term of the likelihood compares the predicted and {\it Gaia}-constrained parameters while accounting for the full covariance matrix; the RV term assumes Gaussian RV uncertainties and that RVs are not covariant with other quantities. Some fitting parameters (e.g., $P_{\rm orb}$ and $e$) are constrained by both RVs and astrometry. Others are constrained only by astrometry (e.g. $\alpha$, $\delta$, $i$) or only by RVs (e.g. $\gamma$). $M_\star$ is constrained only by the prior. We assume broad, flat priors on all parameters except $M_\star$, for which we assume a Gaussian prior $\mathcal{N}(0.79, 0.01)$.  We again use \texttt{emcee} to sample from the posterior, drawing 5000 samples with 64 walkers after a burn-in period of 5000 steps. The results are reported in the ``{\it Gaia}+RVs'' block of Table~\ref{tab:system}.

\subsection{Astrometry alone}
\label{sec:only_astrometry}
We also consider constraints from the {\it Gaia} astrometry alone. In this case, we repeat the procedure described in Section~\ref{sec:joint_fit} but remove the RV term from the likelihood function. Because the binary's center-of-mass RV, $\gamma$, is unconstrained without RVs, we fix it to the value inferred from the joint fit. The resulting constraints are essentially a re-parameterization of the {\it Gaia} constraints into Campbell elements, with the only added information coming from the prior on $M_\star$. These constraints are listed in the ``{\it Gaia} only'' block of Table~\ref{tab:system}.

\begin{figure*}
    \centering
    \includegraphics[width=0.9\textwidth]{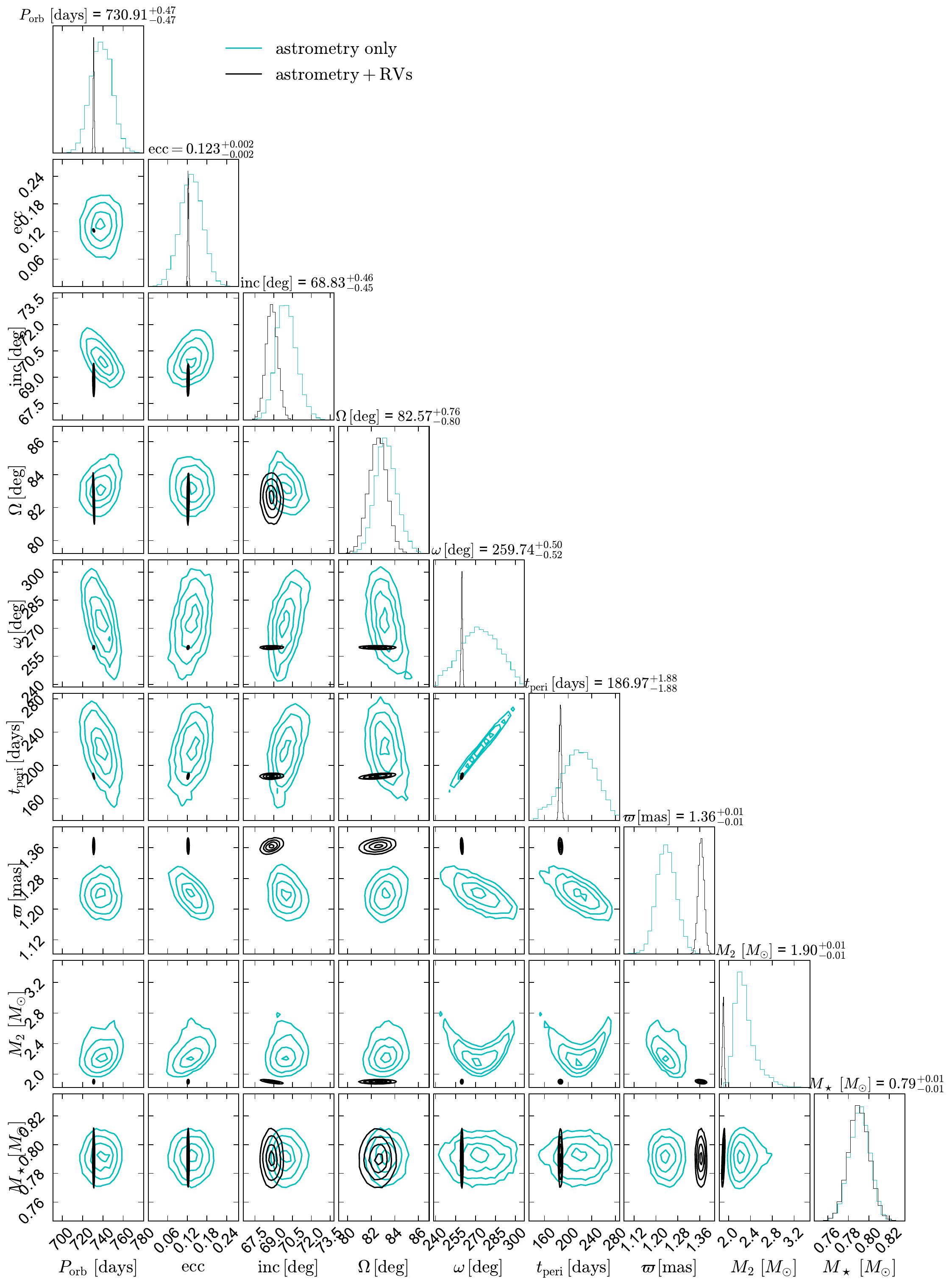}
    \caption{Comparison of constraints from the astrometric solution alone (cyan) and a joint fit of the astrometry and RVs (black). Contours enclose 12\%, 39\%, 68\%, and 86\% probability mass \citep[see][]{corner}.  The constraints from astrometry+RVs are tighter for all parameters except $M_\star$. The two sets of constraints are consistent for most parameters, with the largest tensions being for the parallax $\varpi$ ($3\sigma$ tension) and companion mass $M_2$ ($2\sigma$ tension). The modest tension between the solutions leads us to inflate the astrometric uncertainties in fitting (Section~\ref{sec:inflate_uncert}); the resulting constraints can be found in Figure~\ref{fig:corner_inflate}. }
    \label{fig:corner}
\end{figure*}

\subsection{Predicted RVs}
\label{sec:fitting_results}
Figure~\ref{fig:rvfig} compares the observed RVs to predictions from the fits described above. The top panel shows predictions from the {\it Gaia}-only fit as cyan lines. Individual lines show 50 draws from the posterior, with the spread in these predictions representing the uncertainty in the {\it Gaia} constraints. Despite some uncertainty in the predicted phase at the time of our observations, there is good agreement between the observed and predicted RVs in terms of overall shape, period, and phase. The total variability amplitude of the observed RVs is a few $\rm km\,s^{-1}$ lower than the median value predicted by the {\it Gaia}-only constraints. 

The middle panel of Figure~\ref{fig:inc_fig} compares the observed RVs to predictions from the joint {\it Gaia}+RV fit. These are much more tightly constrained than the {\it Gaia}-only predictions and are in good agreement with the data. The bottom panel shows the residuals between the observed RVs and best-fit joint model. These are consistent with 0 in most cases. The reduced $\chi^2$ is $1.25$, suggesting that the RV uncertainties may be underestimated by $\approx 10\%$ on average.

The relative roles of the RVs and astrometry in constraining the companion mass can be seen in the lower right panel of Figure~\ref{fig:inc_fig}, which shows the companion mass implied by the RV mass function and our constraints on $M_{\star}$ as a function of the assumed inclination. The gray shaded region shows the uncertainty in $M_2$ associated with the uncertainty in both $M_\star$ and the RV mass function. The {\it Gaia} inclination constraint, $i = 68.8\pm 0.5$ deg, implies a unseen companion mass of $M_2=1.902\pm 0.015\,M_{\odot}$. If we jettison the {\it Gaia} solution completely and consider only the constraint from the RV mass function, we obtain a minimum companion mass of $M_2=1.67\pm 0.01\,M_{\odot}$: still $\approx 25\sigma$ above the Chandrasekhar mass.

\subsection{Consistency of the astrometric and RV constraints}
\label{sec:consistency}

Figure~\ref{fig:corner} compares the posterior constraints of key physical parameters obtained from joint fitting of the RVs and {\it Gaia} astrometry (black) and astrometry alone (cyan). Constraints on most parameters are consistent across the two solutions. The parameters not shown in the corner plot due to space constraints ($\alpha$, $\delta$, $\mu_{\alpha}^*$, $\mu_{\delta}$, and $\gamma$) are all consistent at the $1\sigma$ level. The best-fit joint solution also predicts Thiele-Innes coefficients that are in agreement with the {\it Gaia} constraints to $1.1\sigma$ or better. 

The most significant disagreement between the two sets of solutions is in the parallax, $\varpi$, where the disagreement reaches $3\sigma$. This may seem surprising, given that our RVs do not directly constrain the parallax at all! The closer distance of the joint solution can be understood as a result of several compounding factors. First, Figure~\ref{fig:corner} shows that the astrometric solution includes significant negative covariances between $\varpi$ and the parameters $\omega$ and $T_{p}$, which are constrained by RVs. The RVs favor lower values of both parameters than the astrometric solution, drawing the joint fit toward larger $\varpi$. Second, the Thiele-Innes coefficients constrained by {\it Gaia} are functions of the photocenter semi-major axis, $a_0$. The RVs constrain the physical semi-major axis of the luminous star's orbit, $a_\star=a_0/\varpi$, to be smaller than in the best-fit astrometry-only solution. 
This leads to a smaller predicted $a_0$ at fixed distance. The larger parallax favored by the joint fit compensates for this, resulting in a larger $a_0$, which is then compatible with the {\it Gaia} constraints on the Thiele-Innes coefficients.

If we do {\it not} leave the parallax free in the joint fit, we find a higher median companion mass, $M_2 = 2.03\,M_{\odot}$, and a lower inclination, $i = 66.2$ deg.  These covariances can be understood as arising from the fact that the RVs constrain $K_\star$, which is related to $a_0$, $\varpi$, and $i$ through the equation

\begin{equation}
    \label{eq:Kstar}
    K_{\star}=\left(\frac{2\pi a_{0}}{P_{{\rm orb}}\varpi}\right)\frac{\sin i}{\sqrt{1-e^{2}}}. 
\end{equation}
Thus, a larger $\varpi$ or smaller $i$ can both result in a smaller predicted $K_\star$ at fixed $a_0$.

Figure~\ref{fig:rvfig} shows that the observed RVs approximately fall on top of the predictions of the {\it Gaia}-only solution: the main tension is that the predicted RV semi-amplitude is $\approx 2\,\rm km\,s^{-1}$ larger than observed. This suggests that the {\it Gaia} constraints are basically correct; i.e., there was no catastrophic failure in the astrometry. However, it is still possible (and indeed, likely) that the uncertainties of the {\it Gaia}-only solution are underestimated somewhat: one parameter is discrepant between the two fits at the $3\sigma$ level, and our analyses of Gaia BH1 also suggested that the uncertainties of its astrometric solutions were underestimated at the factor of $\sim$2 level \citep{El-Badry2023, Chakrabarti2023, Nagarajan2023}.  

\subsection{Joint fit with inflated astrometric uncertainties}
\label{sec:inflate_uncert}

To assess the effects of possible  unrecognized systematics and/or underestimated uncertainties in the {\it Gaia} solution, we inflated the astrometric solution by a factor of 3 while maintaining the same parameter correlations reported in {\it Gaia} DR3. This is equivalent to multiplying the covariance matrix by 9. We then repeated the joint fit as described in Section~\ref{sec:joint_fit}. The results are reported in the last block of Table~\ref{tab:system}. This inflation makes the astrometry-only and astrometry+RV constraints fully consistent at the $1\sigma$ level. The companion mass constraint then changes to $M_2=1.90\pm 0.04\,M_{\odot}$. We adopt this as our fiducial and most robust value.

The choice of a factor of 3 uncertainty inflation is likely somewhat conservative since it makes {\it all} parameters consistent within $1\sigma$ (see Appendix~\ref{sec:appendix}).

\subsection{Astrometric phase coverage}
\label{sec:phase_coverage}

\begin{figure*}
    \centering
    \includegraphics[width=\textwidth]{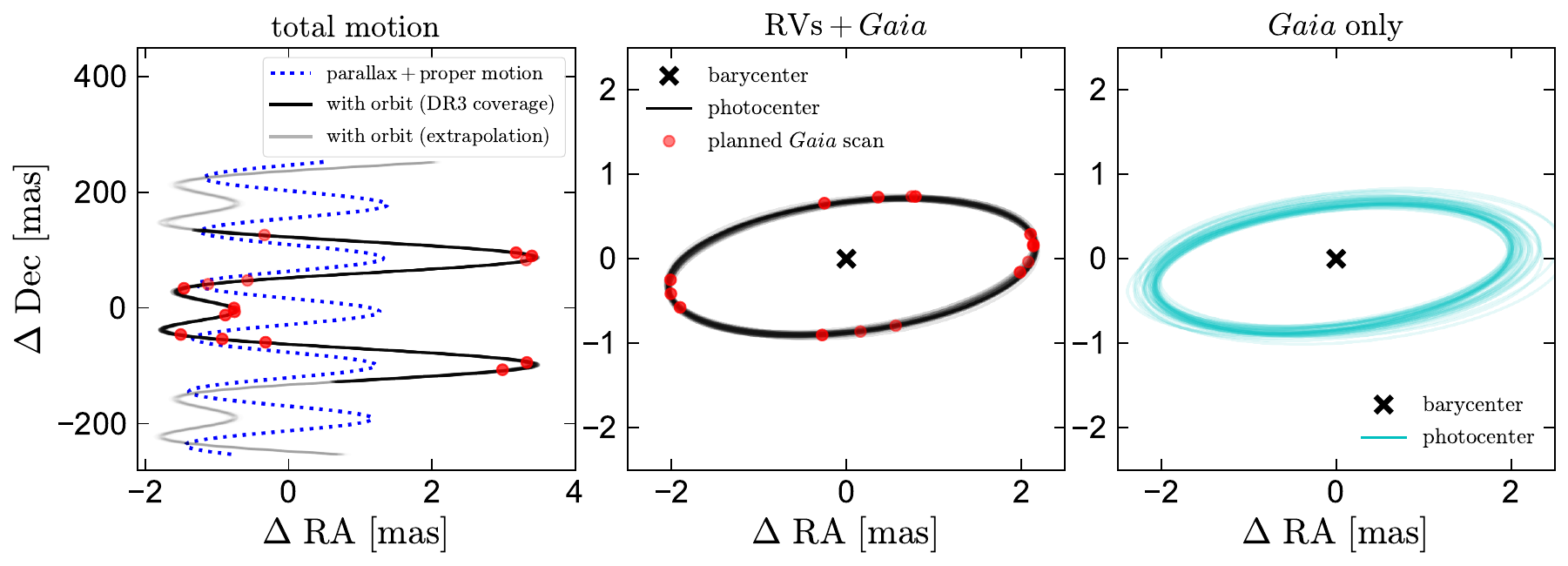}
    \caption{Observation times of \gns predicted by the {\it Gaia} observation scheduling tool (GOST). Black line shows the best-fit orbit from our combined fit. Red points show the predicted photocenter positions at the times when  {\it Gaia} observed the source according to GOST. The actual measured $\Delta \rm RA$ and  $\Delta \rm Dec$ are not published, only the predicted scan times. The expected per-epoch astrometric uncertainties are about 0.2 mas in the along-scan direction. The predicted scan times show that the orbit has been sampled well.}
    \label{fig:scans}
\end{figure*}

Figure~\ref{fig:scans} explores the expected orbital motion and {\it Gaia} observations of Gaia NS1. In the left panel, we show the total plane-of-the-sky motion predicted by the joint RVs+{\it Gaia} solution, separating the effects of parallax and proper motion from those of orbital motion. Note that the axis scales in RA and Dec are very different, because the source's proper motion is almost exactly aligned with the declination axis.\footnote{This apparent coincidence is verified by a proper motion measurement from UCAC5 \citep{Zacharias2017}, which found $\mu_{\alpha}^*=-1.3\pm 2.2\,\rm mas\,yr^{-1}$ and $\mu_{\delta}=-92.0\pm 1.4\,\rm mas\,yr^{-1}$, without accounting for binary motion.} In the middle and right panels, we show the orbital motion only, and compare constraints from the joint fit to those from {\it Gaia} alone. As expected, the joint RV+astrometry solution results in a tighter constraint on the orbital ellipse than the astrometry alone.

The red symbols show times when the source is predicted to have been observed by the {\it Gaia} observation
scheduling tool (GOST).\footnote{\href{https://gaia.esac.esa.int/gost/}{https://gaia.esac.esa.int/gost/}} The actual epoch-level astrometry is not published in DR3, so we simply plot the coordinates predicted by the best-fit joint solution at the timestamps when GOST predicts the source would have transited across the {\it Gaia} focal plane. Due to CCD chip gaps and various issues that can cause gaps in the datastream, only about 90\% of predicted focal plane transits result in a usable astrometric measurement on average. 

For \gn, GOST predicts 26 transits and 15 visibility periods (i.e., 15 groups of observations separated from other groups by at least 4 days), but the \texttt{gaia\_source} table reports that only 24 transits and 14 visibility periods were used in the astrometric solution. This implies that the scan times shown in Figure~\ref{fig:scans} are a serviceable but not perfect approximation of the actual scans used in the astrometric solution. The predicted scan times result in good coverage of the astrometric orbit. No matter which two predicted observations were not used for the astrometric solution, the ellipse would still be reasonably well sampled by any combination of 24 predicted transits in 14 visibility periods. This implies better astrometric phase coverage than was available   for Gaia BH1, where all the predicted astrometric observations fell on one half of the orbit \citep{El-Badry2023}.

\section{Spectroscopic analysis}
\label{sec:spec_anal}
We analyzed the highest-SNR MIKE spectrum of \gns to obtain a spectroscopic estimate of the luminous star's atmospheric parameters and abundance pattern. The spectrum was obtained on JD 2460037.6527 with a 2160 s exposure and has $\rm SNR=102$ at 6500\,\AA. Our analysis follows the hybrid spectroscopic/isochrone fitting method described in \citet{reggiani2022}. In brief, we first infer the effective temperature, $T_{\rm eff}$, surface gravity $\log g$, metallicity, $\rm [Fe/H]$, and microturbulent velocity, $\xi$, through an iterative fit of the broadband SED and the equivalent widths (EWs) of \ion{Fe}{1} and \ion{Fe}{2} atomic absorption lines. We use an iterative approach that minimizes the dependence of the iron abundances inferred from individual lines on reduced equivalent width, while requiring that the atmospheric parameters, parallax, and SED are consistent with MIST isochrones. After the atmospheric parameters have been constrained, we infer individual elemental abundances while fixing the atmospheric parameters to the values measured in the first step.

For the isochrone fitting, we include multiwavelength photometry from the ultraviolet to the near-infrared: GALEX {\it NUV}, Gaia DR2 {\it G}, 2MASS {\it J, H}, and {\it Ks}, and WISE {\it W1} and {\it W2}. We also include our updated parallax ($\varpi=1.36\pm0.03$; Table \ref{tab:system}), and the extinction from the \citet{Green2019} 3D dust map. We use the \texttt{isochrones} package\footnote{\url{https://github.com/timothydmorton/isochrones}} \citep{morton2015} to fit the MESA Isochrones and Stellar Tracks \cite[MIST;][]{Dotter2016,Choi2016} library to the photospheric stellar parameters and multiwavelength photometry, parallax, and extinction data using \texttt{MultiNest}\footnote{\url{https://ccpforge.cse.rl.ac.uk/gf/project/multinest/}} \citep{feroz2008,feroz2009,feroz2019} via \texttt{PyMultinest} \citep{buchner2014}.  We measured EWs using the \textit{Spectroscopy Made Harder} (smh) python package \citep{casey2014}, fitting unblended atomic transitions with Gaussian profiles. The atomic data for each line as well as the measured EWs are available upon request. We assume \citet{asplund2021} photospheric solar abundances. The atmospheric parameters from the hybrid spectroscopic/isochrone fit are listed in Table \ref{tab:system}.

We next inferred elemental abundances of \ion{O}{1}, \ion{Na}{1}, \ion{Mg}{1}, \ion{K}{1}, \ion{Ca}{1}, \ion{Ti}{1}, \ion{Ti}{2}, \ion{Cr}{1}, \ion{Cr}{2}, \ion{Mn}{1}, \ion{Fe}{1}, \ion{Fe}{2}, \ion{Ni}{1}, \ion{Zn}{1}, \ion{Y}{2}, \ion{Zr}{2}, and \ion{Ba}{2} from their EWs, including isotopic/hyperfine splitting details where needed. We again measure the EWs by fitting Gaussian profiles using \textit{smh}. We assume local thermodynamic equilibrium (LTE) and use the 1D plane-parallel, $\alpha$-enhanced \cite{castelli2004} model atmospheres and the 2023 version of the LTE radiative transfer code \texttt{MOOG} \citep{sneden1973} to infer elemental abundances based on our EWs. We report in Table \ref{elem_abundances} our LTE abundance inferences and uncertainties. Using spectral synthesis we also derived the abundances of lithium, cobalt, and europium. The lithium abundance was derived through spectral synthesis of the $6707$ \AA \ transition, for cobalt we used the $3894, \ 3995, \ 4020, \ 4110, \ \rm{and} \ 4118 \ $ \AA \ transitions, and for europium the abundance was derived via the $4129, \ 4205, \rm{and} \ 6645 \ $ \AA \ transitions. All linelists used for the synthesis analyzes are available upon request. For lithium we also performed corrections for 3D, and non-LTE effects using the \textit{breidablik} code \citep{wang2021}.

Our fit yields atmospheric parameters in good agreement with those inferred from the SED fit (Section~\ref{sec:lum_mass}). In most respects, the abundance pattern is unremarkable for an halo star. The $\alpha$ elements O, Ca, Mg, and Ti are enhanced ($[\rm \alpha/Fe]\approx +0.2$) and there is no evidence for s-process enhancement ($[\rm Ba/Fe]= -0.17\pm 0.13$). The most unusual feature of the abundance pattern is a strong enhancement in lithium, $\rm A(Li) = 2.92\pm 0.09$ after the 3D NLTE abundance correction. This value is almost 1 dex higher than typical for stars with temperatures similar to \gn. We discuss the lithium enhancement further in Section~\ref{sec:lithium_galah}.

\begin{table}
\centering
\caption{Elemental Abundances NS1}
\label{elem_abundances}
\begin{tabular}{lrrcrcc}
\hline
Species & $A(\text{X})$ & [X/H] & $\sigma_{\text{[X/H]}}$ & [X/Fe] & $\sigma_{[\text{X/Fe}]}$ &
$n$\\
\hline
\ion{Li}{1}$_{\rm{1D\,LTE}}$ & $3.06$ & $\cdots$ & $\cdots$ & $\cdots$ & $0.1$ & $1$ \\ 
\ion{Li}{1}$_{\rm{3D\,NLTE}}$ & $2.92$ & $\cdots$ & $\cdots$ & $\cdots$ & $0.08$ & $1$ \\ 
\ion{O}{1}  & $7.83$ & $-0.86$ & $0.07$ & $0.37$ & $0.13$ & $3$ \\ 
\ion{Na}{1} & $5.23$ & $-0.99$ & $0.06$ & $0.24$ & $0.20$ & $2$ \\ 
\ion{Mg}{1} & $6.55$ & $-1.00$ & $0.10$ & $0.23$ & $0.13$ & $5$ \\ 
\ion{K}{1}  & $4.49$ & $-0.58$ & $0.00$ & $0.65$ & $0.13$ & $1$ \\ 
\ion{Ca}{1} & $5.30$ & $-1.00$ & $0.13$ & $0.23$ & $0.12$ & $20$ \\ 
\ion{Sc}{1} & $1.86$ & $-1.28$ & $0.09$ & $-0.05$ & $0.11$ & $4$ \\ 
\ion{Ti}{1} & $3.95$ & $-1.02$ & $0.17$ & $0.21$ & $0.17$ & $9$ \\ 
\ion{Ti}{1} & $4.03$ & $-0.94$ & $0.16$ & $0.29$ & $0.09$ & $19$ \\ 
\ion{Cr}{1} & $4.33$ & $-1.29$ & $0.11$ & $-0.06$ & $0.18$ & $7$ \\ 
\ion{Mn}{1} & $3.79$ & $-1.63$ & $0.00$ & $-0.40$ & $0.11$ & $1$ \\ 
\ion{Fe}{1} & $6.28$ & $-1.18$ & $0.10$ & $0.05$ & $0.12$ & $36$ \\ 
\ion{Fe}{1} & $6.20$ & $-1.26$ & $0.10$ & $-0.03$ & $0.08$ & $14$ \\ 
\ion{Co}{1} & $3.59$ & $-1.35$ & $0.10$ & $-0.12$ & $0.12$ & $5$ \\
\ion{Ni}{1} & $5.04$ & $-1.16$ & $0.21$ & $0.07$ & $0.22$ & $9$ \\ 
\ion{Zn}{1} & $3.31$ & $-1.25$ & $0.10$ & $-0.02$ & $0.10$ & $3$ \\ 
\ion{Y}{1}  & $0.94$ & $-1.27$ & $0.32$ & $-0.04$ & $0.21$ & $5$ \\ 
\ion{Zr}{1} & $1.65$ & $-0.94$ & $0.22$ & $0.28$ & $0.18$ & $3$ \\ 
\ion{Ba}{1} & $1.12$ & $-1.15$ & $0.08$ & $0.08$ & $0.15$ & $3$ \\ 
\ion{Eu}{2} & $-0.62$ & $-1.14$ & $0.10$ & $0.09$ & $0.15$ & $3$ \\
\hline
\end{tabular}
\label{tab:abund}
\end{table}

\subsection{Comparison of the spectrum to similar stars}
\label{sec:galah}
\begin{figure*}
    \centering
    \includegraphics[width=\textwidth]{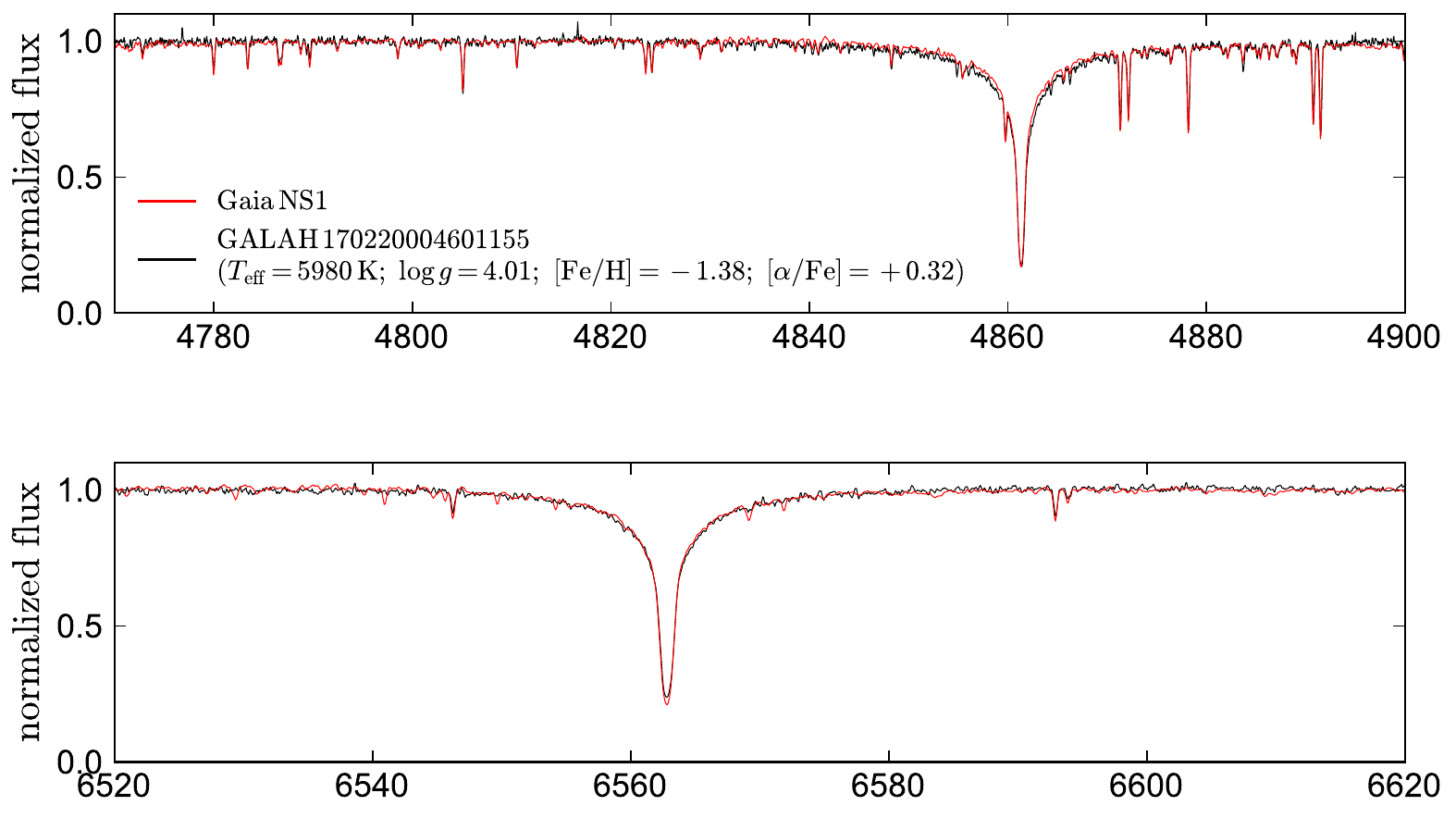}
    \caption{Spectral cutouts of \gns (red) compared to a reference star (black), a halo star with similar stellar parameters and abundances observed by the GALAH survey. The two spectra are very similar, ruling out significant light contributions from a companion and validating the stellar parameters and abundances inferred in Section~\ref{sec:spec_anal}. }
    \label{fig:galah}
\end{figure*}

To search for possible anomalous spectral features, we compared the spectrum of \gns to spectra of stars with similar stellar parameters and abundances observed by the GALAH survey \citep{DeSilva2015, Buder2021}. To this end, we degraded the highest-SNR MIKE spectrum to $R=28,000$, shifted it to rest frame, and identified its nearest neighbor in pixel space among stars observed by GALAH with SNR $>$ 50. The closest match was {\it Gaia} DR3 source 6251344007644295680 (GALAH DR3 object ID 170220004601155), which has $T_{\rm eff} = 5980\pm 78\,\rm K$, $\log g = 4.01\pm 0.18$, $\rm [Fe/H]=-1.38\pm 0.08$, and $[\rm \alpha/Fe]=0.32\pm 0.02$. These parameters are all similar to those we find for \gn, providing independent validation of our inferred parameters. In Figure~\ref{fig:galah}, we compare the normalized and resolution-matched spectra of the two sources in the 1st and 3rd GALAH wavelength windows, which respectively contain H$\beta$ and H$\alpha$. The spectra are very similar. The most obvious differences are in the wings of the H$\beta$ line -- which could reflect either differences in continuum normalization or slight differences in $\log g$ -- and in small telluric absorption lines near the H$\alpha$ line. The two sources also have identical dereddened $G_{\rm BP}-G_{\rm RP}$ colors within 0.01 mag, and identical absolute magnitudes within 0.03 mag when we assume our fiducial parallax of $\varpi = 1.36$\,mas for \gn. Like \gn, the star is on a halo orbit, with a plane-of-the-sky tangential velocity $v_{\perp} \approx 370\,\rm km\,s^{-1}$. The similarity of the two observed spectra and the sources' absolute magnitudes both speaks against the presence of a luminous secondary and validates the revised parallax from our joint {\it Gaia}+RV fit. 

\subsection{Lithium enhancement}
\label{sec:lithium_galah}
\begin{figure*}
    \centering
    \includegraphics[width=\textwidth]{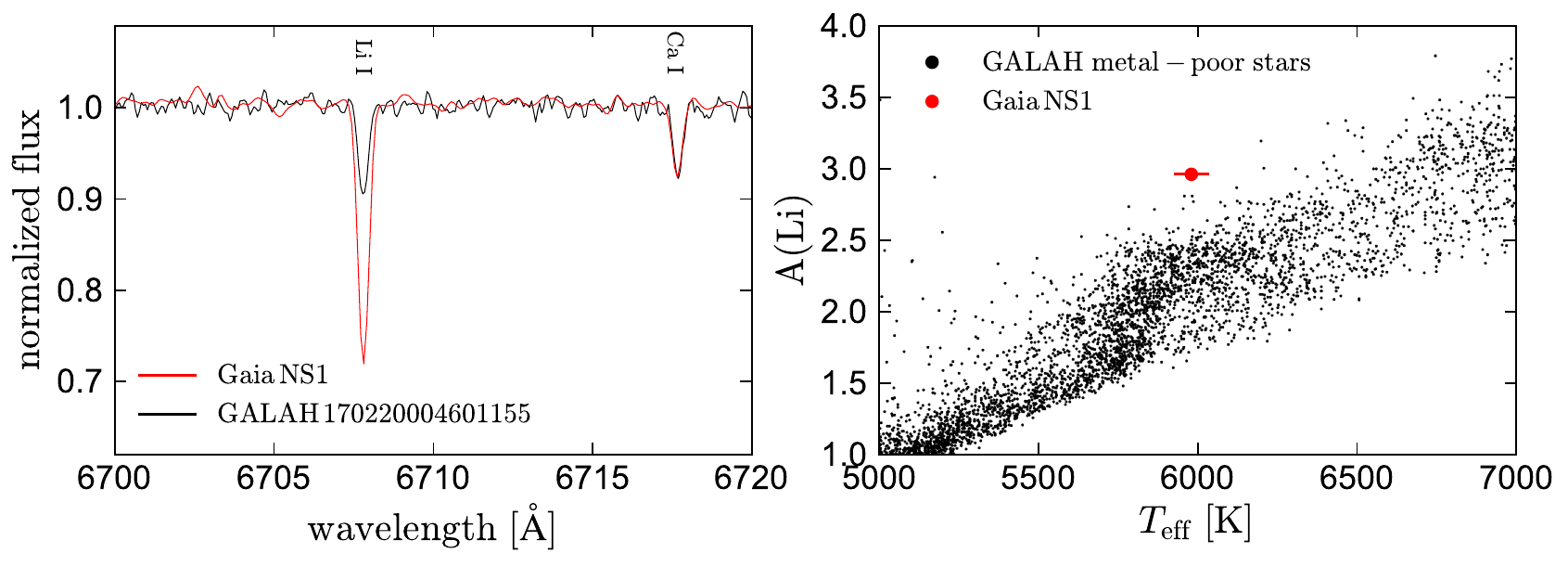}
    \caption{Left: observed spectrum of \gns (red) compared to the reference star from the GALAH survey whose spectrum most closely matches \gns (Figure~\ref{fig:galah}). \gns is strongly enhanced in lithium compared to the reference star. Right: lithium abundance of \gns compared to stars observed by GALAH with $\rm [Fe/H] < -0.7$ and $\rm SNR > 30$. \gns is clearly an outlier in lithium abundance: it has higher A(Li) than any of the 1000+ stars with $T_{\rm eff}$ within $\pm 200$ K.}
    \label{fig:galah_lithium}
\end{figure*}

Although most of the spectrum of \gns is very similar to the metal-poor turnoff stars observed by GALAH, there is one respect in which the object is unique: it has an unusually strong Li I\,$\lambda6708$ line, indicative of a high lithium abundance. This is illustrated in Figure~\ref{fig:galah_lithium}. The equivalent width of the line is 114\,m\AA, 
more than 3 times larger than the median value for $\approx 35$\,m\AA\,for the top 20 closest spectral doppelgangers. Our inferred lithium abundance, $\rm A(Li) = 2.92\pm 0.08$, is higher than that of {\it any} star observed by GALAH with similar parameters and a high-SNR spectrum (right panel of Figure~\ref{fig:galah_lithium}). Within a temperature range of $\pm 200$\,K and a metallicity range $[\rm Fe/H] < -0.7$, there are 1097 stars observed by GALAH with $\rm SNR > 30$, and none has lithium enhancement comparable to \gn. We inspected the spectra of the most lithium-rich GALAH sources with temperatures comparable to \gns and found most to have spectral artifacts near the Li I\,$\lambda6708$ line, pointing to spurious $\rm A(Li)$ measurements. We did not find {\it any} GALAH sources with similar parameters to \gns and comparably strong Li I\,$\lambda6708$ lines. The lithium enhancement of \gns is thus very significant.

Lithium enhancement is commonly associated with young stars, since lithium is destroyed in stellar interiors. \gns is obviously not young given its low metallicity and CMD position near the main-sequence turnoff. Strong lithium enhancement has also been found in the donors of several BH and NS X-ray binaries \citep[e.g.,][]{Martin1992, Marsh1994, Martin1994, Filippenko1995, Martin1996}. A clear reason for this enhancement has never been identified, with proposed explanations including lithium production in supernovae \citep[e.g.,][]{Dearborn1989}, spallation on the surface of the companions due to the neutron flux from the compact object's accretion flow \citep[e.g.,][]{Guessoum1999, Fujimoto2008}, spallation within the accretion flow and subsequent pollution of the companion by winds \citep{Yi1997}, and slowing of lithium destruction in the companions due to their rapid rotation \citep{Maccarone2005}. In \gn, the wide separation of the star and companion, negligible expected accretion rate of the compact object, and slow rotation of the companion make all these proposed explanations seem unlikely.

Lithium is also produced in AGB and super-AGB stars \citep[e.g.,][]{Cameron1971, Sackmann1992, Ventura2010}, raising the question of whether the companion could be a WD. Our measured companion mass is far too high for it to be a single WD, but WD+WD binary models are in principle possible (Section~\ref{sec:companion_nature}). Lithium enhancement is not, however, commonly observed in stars with WD companions and other abundance anomalies associated with pollution from AGB companions \citep{Pinsonneault1984, Allen2006}. Moreover, most lithium-enhanced stars do not exhibit significant RV variability \citep{Castro-Tapia2023, Sayeed2023}.  We have not found evidence of lithium enhancement in our follow-up of astrometric binaries from {\it Gaia} with inferred masses significantly below the Chandrasekhar mass.   

The other two metal-poor NS candidates we have identified (El-Badry et al.~2024, in prep) {\it also} have very strong lithium enhancement. Compared to \gn, the other two candidates with $\rm [Fe/H] < -1$ have lower companion masses ($M_2 \approx 1.30\,M_\odot$ and $\approx 1.42\,M_{\odot}$) and more eccentric orbits ($e=0.66$ and $e=0.68$). These masses are typical of normal NSs, and the higher eccentricities are consistent with what is expected from natal kicks.  We have found little evidence for lithium enhancement in any of the candidates with metallicities near solar. 

Assuming the enhancement is due to some form of pollution from a companion, it is expected that it should be stronger at low metallicity: low-metallicity stars have thinner convective envelopes, meaning that the enriched material is mixed with less material from the luminous star. This is a fairly significant effect: from MESA models of $0.8\,M_{\odot}$ stars, we find that in a model with initial metallicity $Z=0.0014$, the convective envelope contains only $\approx 0.2$\% of the star's mass at the observed evolutionary phase of \gn, compared to $\approx 10\%$ in a model with $Z=0.014$. That is, accreted material is diluted $\sim 50$ times less on the surface of the luminous star in \gns than it would be in a solar-metallicity companion.

\section{Galactic orbit}
\label{sec:galpy}

\begin{figure*}
    \centering
    \includegraphics[width=\textwidth]{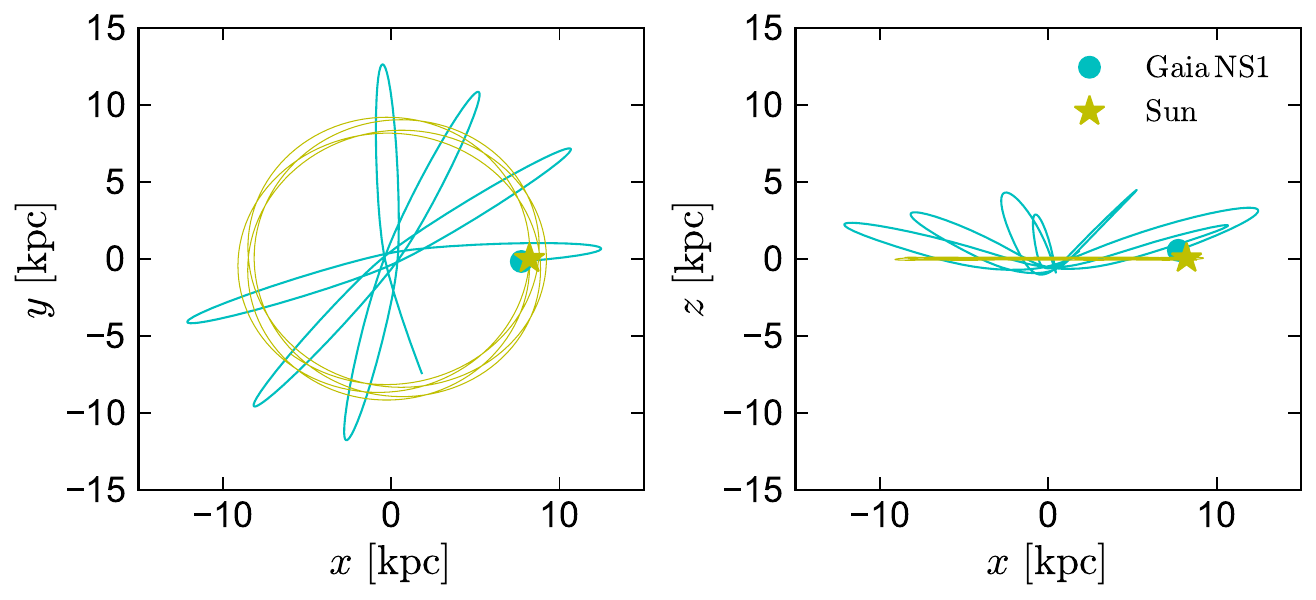}
    \caption{Galactic orbit of \gn, integrated backwards for 1\,Gyr. The orbit of the Sun is shown for comparison. \gns has an eccentric orbit that travels several kpc from the Galactic disk, characteristic of a halo star.  }
    \label{fig:galpy}
\end{figure*}

To investigate the Galactic orbit of \gn, we used the parallax, proper motion, and center-of-mass RV from the joint {\it Gaia}+RV fit as starting points to compute its orbit backward in time for 1 Gyr using \texttt{galpy} \citep[][]{Bovy2015}. We used the Milky Way potential from \citet{McMillan2017}. The system's current LSR-corrected Galactic space velocity is $\rm UVW=(218, -250, -88)\,\rm km\,s^{-1}$.

The orbit is shown in Figure~\ref{fig:galpy}; for comparison, we also show the orbit of the Sun. The orbit is typical of a halo star: it is highly eccentric and reaches several kpc above the disk midplane. This conclusion is insensitive to variations of the source's astrometric parameters within their uncertainties and/or the adopted Galactic potential. 

We note that the wide orbit and low eccentricity of \gns rule out a large natal kick to the NS. The system's large space velocity must thus be a consequence of its being very old. 

\section{Discussion}
\label{sec:discussion_kicks}

\begin{figure}
    \centering
    \includegraphics[width=\columnwidth]{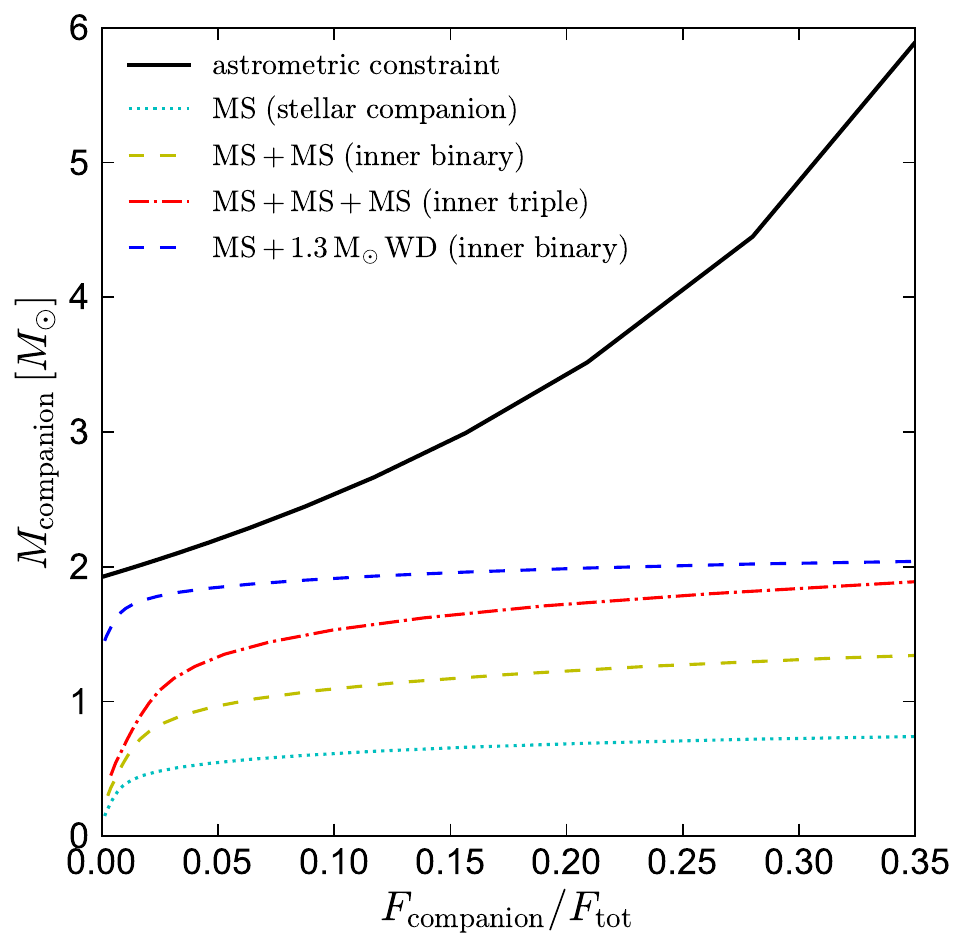}
    \caption{Constraints on the mass of the unseen companion in \gns as a function of the $G-$band flux ratio. Solid black line shows the constraint from the astrometric mass function, assuming $M_\star = 0.79\,M_{\odot}$; a completely dark companion would imply $M_{\rm companion}\approx 1.90\,M_{\odot}$. If the companion contributes some light, its astrometrically-implied mass increases. Dotted cyan line shows the expected flux ratio and mass for a single main-sequence companion. Because this is always below the black line, no single MS companion can explain the orbit. The same is true for an equal-mass inner binary (yellow dashed) or an equal-mass inner triple (red dot-dashed). Even an inner binary containing a $1.3\,M_{\odot}$ WD and a MS star (blue dashed) line has a mass-to-light ratio that is too low to match the observed mass function. This leaves a single NS or a WD-WD binary as the only viable companions.}
    \label{fig:scenarios}
\end{figure}

\subsection{Could the luminous star have an unusually low mass?}
\label{sec:lower_mass}
Several dormant BH candidates published in recent years turned out to be mass transfer binaries in which the luminous star had a much lower mass than assumed when its temperature and radius were compared to single-star models \citep[e.g.,][]{Shenar2020, Bodensteiner2020, El-Badry2021, El-Badry2022_1850}. It is worth considering whether the luminous star in \gns could similarly be a product of mass transfer, in which case the companion's mass could be overestimated as well.


To test whether a pre-WD could conceivable masquerade as a main-sequence star similar to the luminous star in \gn, we calculated a grid of MESA binary evolution models \citep{Paxton2011, Paxton2015}. We considered donor stars with initial masses of $(1-2)\,M_{\odot}$ and initial periods of 50--150\,d, modeling the companion as a $(1-2)\,M_{\odot}$ point mass.  Several of these models detach from their Roche lobes at periods near 730\,d, contract, and become helium WDs. However, their temperatures and radii never simultaneously approach those of main-sequence stars similar to \gn: when they have $R\sim 1\,R_{\odot}$, these models have $T_{\rm eff} = 30,000-40,000$\,K. By the time they cool to $T_{\rm eff}\sim 6000$\,K, they are on the WD cooling track, with $R\sim 0.01\,R_{\odot}$. We thus conclude that stripped stars are unlikely to masquerade as solar-mass main sequence stars.

Moreover, for a pre-WD formed by stable mass transfer, the expected mass at a period of 730\,d is $\approx 0.45\,M_{\odot}$ \citep[e.g.,][]{Rappaport1995}. If the mass of the luminous star were somehow this low, the dynamically-required mass of the companion would still be $M_2\approx 1.55\,M_{\odot}$. This is significantly above the Chandrasekhar limit and would thus most simply imply a NS companion.

At short periods, ``semi-stripped'' donors with thick hydrogen envelopes {\it can} have radii and temperatures similar to solar-mass main sequence stars \citep[e.g.,][]{El-Badry2021_lamost, El-Badry2021_elms}, but the evolutionary scenario that gives rise to such donors requires stable mass transfer to a close companion ($P_{\rm orb} < 1$\,d), which is clearly not applicable here.

\subsection{Nature of the companion}
\label{sec:companion_nature}
All our evidence of the companion comes from its gravitational effects on the luminous star, whose motion we can probe with both astrometry and RV measurements. While we can constrain the total mass of the companion dynamically, this provides only indirect information about its nature. We discuss several possibilities below. 

\begin{enumerate}
\item {\it A single neutron star:} This is the most straightforward explanation and our default assumption in most of the text. 
\item {\it A low-mass black hole:} \gns was initially presented by \citet{Andrews2022} and \citet{Shahaf2023} as a BH candidate, when its inferred mass was $\approx (2.5-2.7)\,M_{\odot}$. Given our lower mass constraint of $\approx 1.90\,M_\odot$, a BH seems unlikely: there are no well-understood astrophysical channels to produce BHs of this mass, and several objects that are definitely NSs have fairly well-constrained masses above $1.9\,M_{\odot}$ \citep[e.g.,][]{Antoniadis2013, Cromartie2020}.

\item {\it One, two, or three main-sequence stars, or other kinds of dim but luminous companions}: A luminous companion that is a single star, a tight main-sequence binary, or even a tight main-sequence triple, is ruled out. This is because any companion that is not dark would require the luminous star's orbit, $a_\star$, to be larger than the photocenter orbit, $a_0$, and thus would require an even larger companion mass. This is illustrated in Figure~\ref{fig:scenarios}, which compares the mass-to-light ratios of various types of companions (dashed and dotted colored lines) to the dynamical constraint from the astrometric mass function (black line). For example, if a hypothetical dim companion contributed 15\% of the $G$--band light, then the companion mass implied by the astrometric mass function would rise to $3.0\,M_{\odot}$. Figure~\ref{fig:scenarios} shows that a WD+MS inner binary is ruled out for the same reason. 

Joint fitting of astrometry and RVs directly constrains the fraction of the light coming from the companion, because a luminous companion implies larger RV variability amplitude at fixed $a_0$. To quantify this limit, we tried fitting $\epsilon$, the $G$-band flux ratio of the companion to the primary, as a free parameter \citep[see][]{El-Badry2023_bh2}. We conservatively inflated the astrometric uncertainties by a factor of 3 (see Section~\ref{sec:inflate_uncert}) and found 1 and $2\sigma$ upper limits of $\epsilon < 0.02$ and $\epsilon < 0.03$, respectively. These limits are quite stringent and rule out almost all scenarios involving non-degenerate companions. 

\item {\it A neutron star + white dwarf binary:} The unseen object could also be a close binary containing a neutron star and a white dwarf, with masses of, e.g., $1.4\,M_{\odot}+0.5\,M_{\odot}$ or $1.7\,M_{\odot}+0.2\,M_{\odot}$. Such an inner binary could be effectively invisible in the optical. How such a system could have formed is quite uncertain. In the $1.7\,M_{\odot}+0.2\,M_{\odot}$ scenario, the inner binary could be a result of stable mass transfer and the NS should be a recycled millisecond pulsar. If the inner binary was very tight, it may subsequently have evolved through an ultra-compact X-ray binary phase, possibly leaving behind an isolated NS \citep[][]{Tauris2023}. In the  $1.4\,M_{\odot}+0.5\,M_{\odot}$ scenario, it would have to be a product of common envelope evolution to fit inside the luminous star's orbit today \citep[e.g.,][]{Rappaport1995}. Either scenario requires a common envelope episode prior to the formation of the NS. This seems difficult to achieve, because a low-mass companion is only expected to survive such an interaction with a red supergiant progenitor if it begins in an orbit of several au, beyond the current orbit of the luminous star \citep[e.g.,][]{Dewi2000}.

There is one known NS in a triple: the millisecond pulsar PSR J0337+1715 \citep{Ransom2014, Kaplan2014}. That system contains a $1.43\,M_{\odot}$ NS, in a 1.6 d orbit with a $0.20\,M_{\odot}$ WD, all orbited by a $0.41\,M_{\odot}$ WD in a 327 d orbit. How that system formed is not well understood: two intriguing but yet uncertain scenarios involving a double common envelope have been proposed \citep{Tauris2014, Sabach2015}. In any case, it is plausible that \gns could have formed through a similar pathway to that system. This could involve, for example, the luminous star being born in a wider orbit than it inhabits today and spiraling inward following interaction with the NS progenitor's unbound envelope after its ejection through a common envelope interaction with an inner star \citep[e.g.,][]{Tauris2014}. 

\item {\it A white dwarf + white dwarf binary:} Our dynamically-inferred mass is also consistent with two WDs. Both would have to be rather massive, e.g., $1.0\,M_{\odot}+0.9\,M_{\odot}$ or $1.3\,M_{\odot}+0.6\,M_{\odot}$. No close WD+WD binaries are known with total mass exceeding $\sim 1.4\,M_{\odot}$ \citep[e.g.,][]{Napiwotzki2020}, so such a binary would be the most massive WD+WD binary known by a large margin. Given that \gns is $\gtrsim 12$\,Gyr old and the WD progenitors would have had MS lifetimes of at most a few Gyr, both WDs would be ancient, colder than the luminous star, and essentially impossible to detect. As with a NS+WD binary, the main weakness of the WD+WD binary scenario is the difficulty of forming such a massive inner binary while retaining the luminous star as a bound tertiary in a stable orbit. Population synthesis simulations of triples with inner WD+WD binaries predict the tightest triples to have outer semimajor axes  $a_{\rm outer}\approx 50$\,au, much wider than \gns \citep{Shariat2023}. We discuss some possible evolutionary scenarios below.  

 The simplest WD+WD binary model is one in which the binary was born tight and the luminous star never interacted with the progenitor of either WD. Such a scenario, is, however, difficult to make work in practice.  Most formation channels for massive WD+WD binaries require the binary to expand to dimensions of order 1 au \citep[e.g.,][]{Ruiter2013}, since tighter orbits lead to early envelope stripping and formation of lower-mass WDs. The total initial mass of the inner binary would have to have been significantly greater than the mass of the dark companion today, since $(0.9-1.0)\,M_{\odot}$ WDs are formed from stars with initial masses of $(4-6)\,M_{\odot}$ \citep[e.g.,][]{ElBadry2018, Cunningham2024}. For a conservative total initial inner binary mass of $8\,M_{\odot}$, the initial orbital period of the luminous star would have been $\approx 69$\,d \citep{Jeans1924}. In order to remain stable, the inner binary period then could not have exceeded $\approx 69/5$\,d during its evolution \citep[e.g.,][]{Mardling2001}. We experimented with a variety of MESA binary models in search of an evolutionary scenario in which a binary remains tight and produces two WDs with total mass $\sim 1.9\,M_{\odot}$ but were unable to find one:  mass transfer widens the orbit too much, unless the initial mass ratio is highly unequal. Binaries with highly unequal initial mass ratios can remain tight, but in these cases, mass transfer becomes dynamically unstable and likely leads to a stellar merger. 

 A more speculative scenario that might lead to a WD+WD inner binary places the initial orbit of the luminous star {\it wider} than its current orbit, e.g. with an initial period of $\sim 10$\,yr. In this case, the (hypothetical) third star could have been born with an initial period of order 1\,yr, gone through a common envelope when the primary became an AGB star, and gone through another phase of interaction when it terminated its main-sequence evolution. The tertiary could, perhaps, spiral inward due to interaction with the AGB star's wind or common envelope ejected preferentially in the orbital plane \citep[e.g.][]{Bodenheimer1984, Passy2012}. An attractive feature of such a scenario -- which is similar to the one explored by \citet{Tauris2014} for PSR J0337+1715 -- is that the tertiary could accrete Li-enriched material from the AGB star's wind. However, it is quite uncertain whether this scenario can work in practice, because the presence of the inner companion will likely suppress mass transfer to the outer tertiary, preventing its inspiral.

\end{enumerate}

We conclude that a single massive NS is the simplest explanation for the data. A single WD is untenable. Scenarios involving an inner binary (WD+WD or WD+NS) cannot be ruled out based on our data alone, but are challenging to explain with evolutionary models. Such scenarios  could be tested with high-precision RV observations \citep[e.g.,][]{Nagarajan2023}, which would be sensitive to inner binaries with periods longer than about a week. The NS+WD scenario can potentially be tested with radio observations, since the NS would likely be a recycled pulsar. However, the beaming fraction even of a millisecond pulsar is only $\sim 0.3$ \citep{Lorimer2012}, so it may not beam in our direction in any case. In the following discussion, we focus on the scenario in which the companion is a single NS. 

\begin{figure*}
    \centering
    \includegraphics[width=\textwidth]{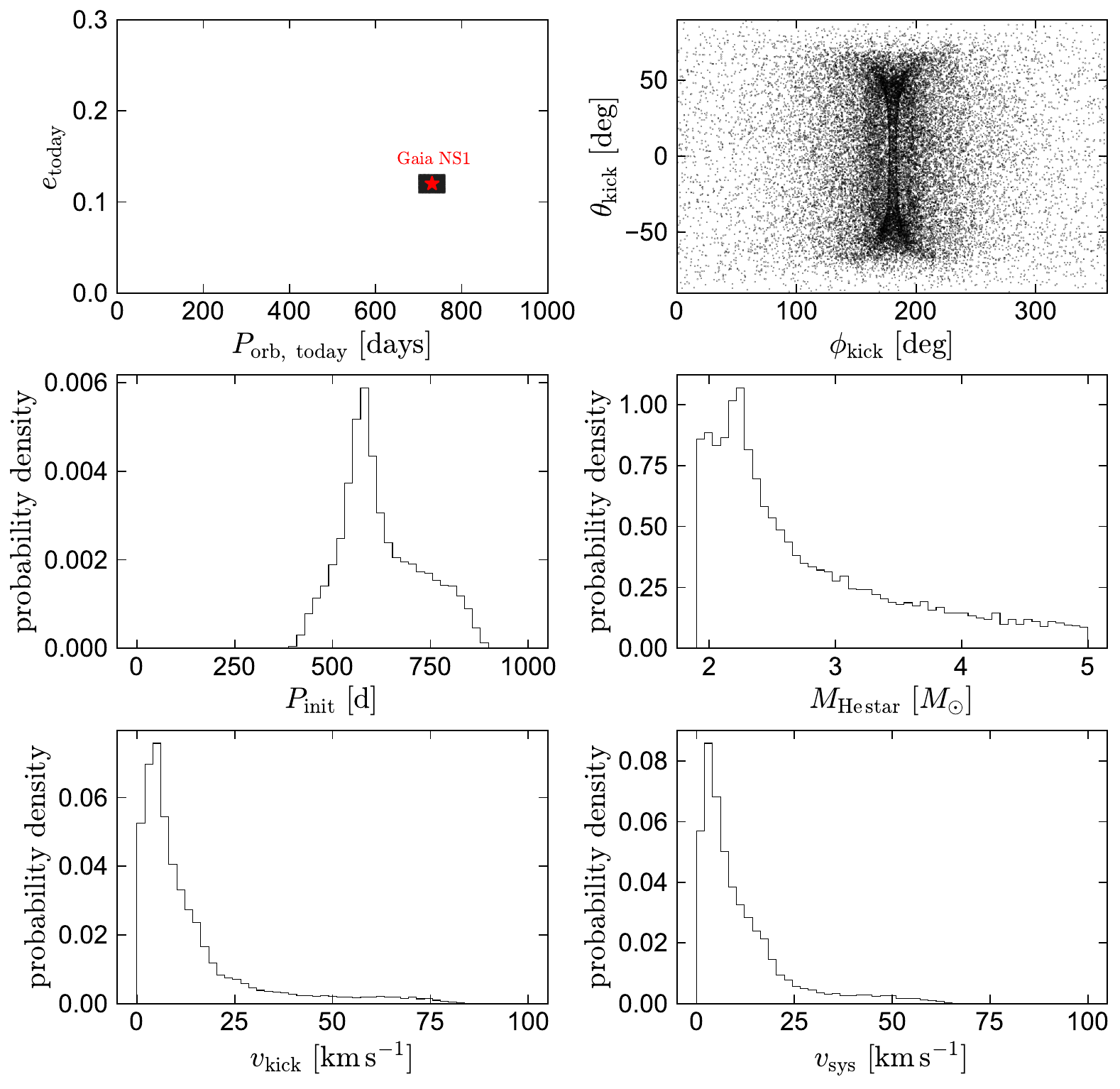}
    \caption{Constraints on kicks and the pre-SN orbit of \gn. We assume that the binary had a circular orbit before the SN and that the NS received a kick with velocity $v_{\rm kick}$ when it formed. We simulate $10^8$ pre-SN orbits with a uniform period distribution $P_{\rm orb}/{\rm d} \sim \mathcal{U}(0,1000)$, a pre-SN mass distribution $M_{\rm He\,star}/M_\odot \sim \mathcal{U}(1.9,5.0)$, and a kick velocity distribution $v_{\rm kick}/({\rm km\,s^{-1}}) \sim \mathcal{U}(0,100)$. We then show properties of the systems whose post-SN orbits have periods and eccentricities similar to \gns (upper left). Most simulated systems that can produce the observed orbit have weak kicks ($v_{\rm kick}\lesssim 20\,{\rm km\,s^{-1}}$) and formed from low-mass He stars ($M_{\rm He\,star}\lesssim 3\,M_\odot$), though there is a long tail of higher kick velocities and more massive He stars that could produce the observed orbit for fine-tuned kick orientations.}
    \label{fig:kickplots}
\end{figure*}

\begin{figure}
    \centering
    \includegraphics[width=\columnwidth]{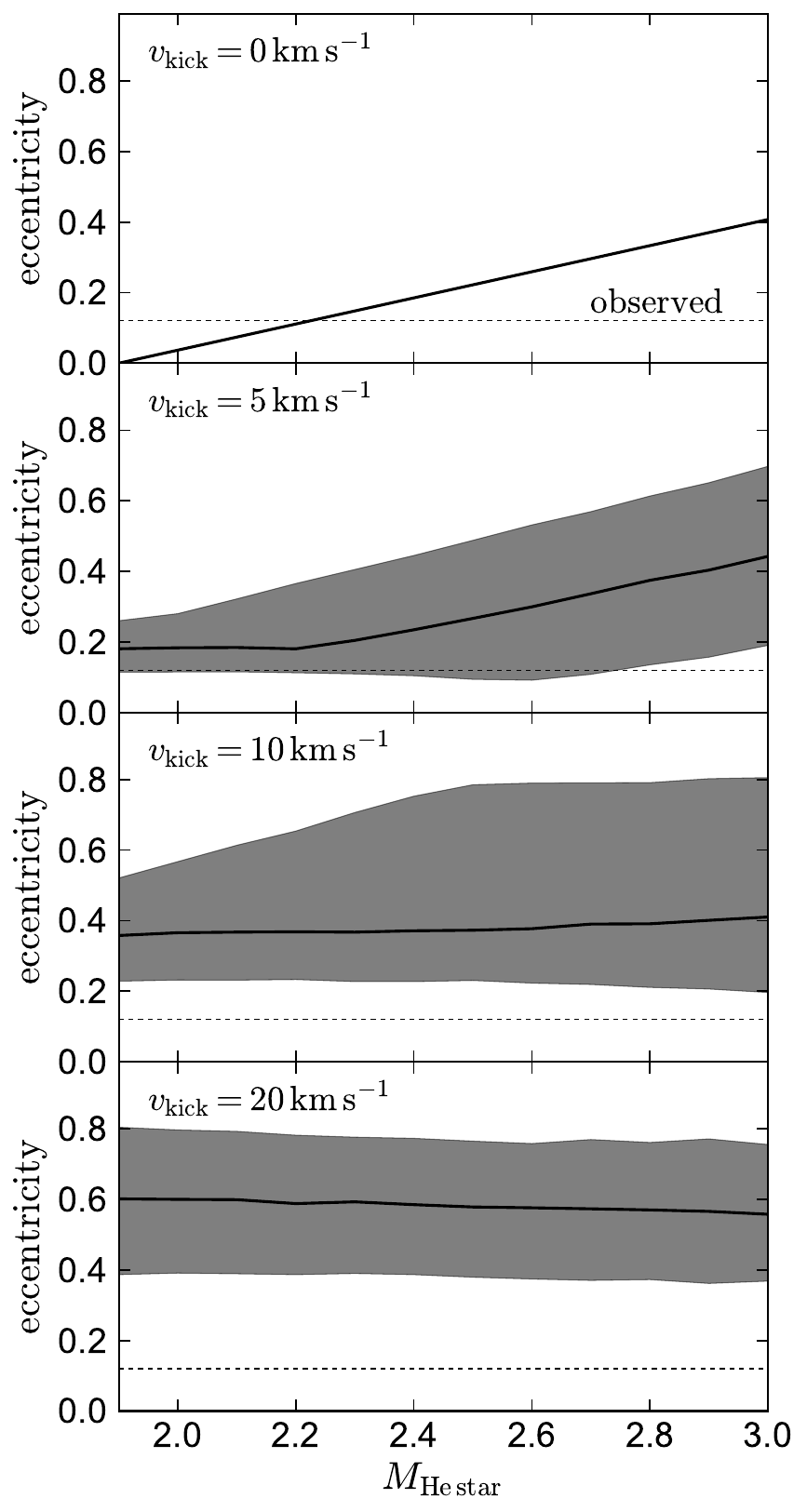}
    \caption{Predicted eccentricity of a $1.9\,M_{\odot}$\,NS + $0.8\,M_{\odot}$\,MS binary following a supernova. We assume the orbit is initially circular, that the NS forms from a helium star of mass $M_{\rm He\,star}$, and that it receives a natal kick of velocity $v_{\rm kick}$ in a random direction at formation. Solid lines and shaded regions show median and middle 68\% eccentricity regions of the surviving orbits. In the absence of a natal kick, loss of more than $0.3\,M_{\odot}$ leads to a higher eccentricity than observed. Kicks generally increase the eccentricity. The low observed eccentricity of \gns thus suggests that the NS formed from a low-mass He star with a small natal kick.   }
    \label{fig:kick}
\end{figure}

\begin{figure}
    \centering
    \includegraphics[width=\columnwidth]{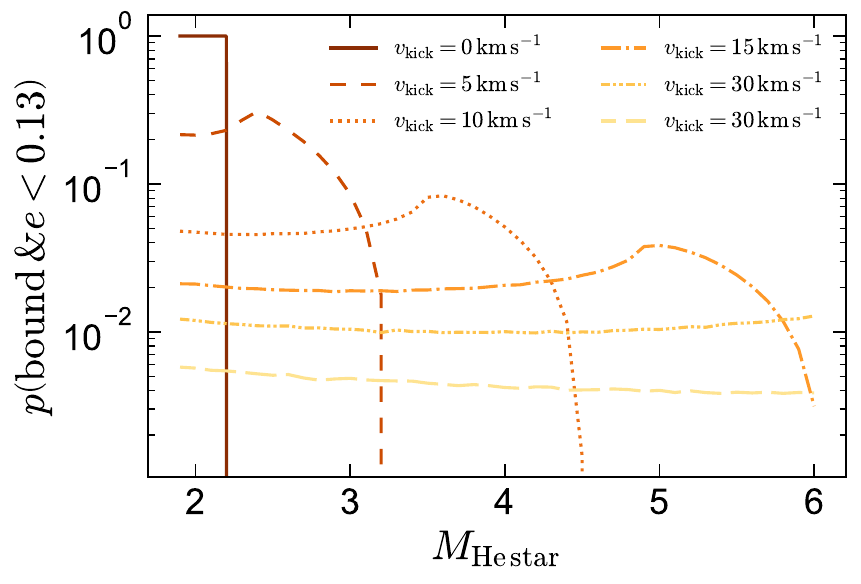}
    \caption{Probability of remaining in a bound orbit with eccentricity $e<0.13$, the value observed for \gn. As in Figure~\ref{fig:kick}, we assume that the orbit is initially circular, that the NS forms from a helium star of mass $M_{\rm He\,star}$, and that it receives a natal kick of velocity $v_{\rm kick}$ in a random direction at formation. For very low $v_{\rm kick}$ values, a low He star mass is required, because mass loss will otherwise result in an eccentric orbit. For higher $v_{\rm kick}$, a bound and near-circular orbit can be maintained if the kick is fortuitously aligned with the companion's instantaneous velocity vector, but the probability of this occurring is low. }
    \label{fig:low_ecc_prob}
\end{figure}

\subsection{Difficulty of explaining the current wide orbit}
\label{sec:wide_orbit}
The current separation between the luminous star and dark companion is $\approx 2.2$\,au. This separation is uncomfortable because it is both too tight for the progenitor of the NS to have fit inside as a red supergiant or AGB star, and too wide to be a result of common envelope evolution. Essentially the same problem applies to the orbits of Gaia BH1 and BH2, to the 20 other NS candidates we have followed-up (El-Badry et al. 2024, in prep.), and perhaps also to most of the $\sim$3,000 WD+MS binaries with astrometric orbits in {\it Gaia} DR3 \citep{Yamaguchi2023, Shahaf2023b}. It is possible that these systems can form through common envelope evolution if the compact object progenitor's envelope is very weakly bound and/or if significant energy is released through recombination  \citep{Webbink2008, Izzard2018, Yamaguchi2023, Belloni2024}. Alternatively, mass transfer may remain stable even under extreme mass ratios \citep[e.g.,][]{Nelemans2000}. The fact that a rather large population of such wide post-interaction binaries exists suggests that their formation does not require exotic conditions or fine-tuned scenarios. The fact that most of the wide WD binaries have near-circular orbits (as does \gn) suggests that these systems are not primarily assembled dynamically, as has been proposed for the BH systems \citep{Rastello2023, DiCarlo2023, Tanikawa2024}.

\subsubsection{Effects of low metallicity}
\label{sec:low_feh} 
We next consider whether the low metallicity of the \gns system could have resulted in a compact NS progenitor that never overflowed its Roche lobe. 
At the current orbital separation of $a\sim 2.2$\,au, a $(10-20)\,M_{\odot}$ red supergiant would overflow its Roche lobe when its radius reached $(280-310)\,R_{\odot}$. Observed red supergiants reach maximum radii of $\sim 1000\,R_{\odot}$ \citep{Yang2020} at SMC metallicity, which is about twice the metallicity of \gn. Observations of massive stars with $Z\lesssim 0.1\,Z_{\odot}$  are scarce \citep[e.g.,][]{Gull2022}, making their late-stage evolution uncertain.

A realistic constraint on the maximum radius of the progenitor if there was no Roche lobe overflow is  more restrictive than the current orbit, because the progenitor would have been more massive than the NS and the orbit would have expanded in response to mass loss. An initial mass of $(10-20)\,M_{\odot}$ implies an initial separation of $(0.3-0.6)$\,au, which would lead to Roche lobe overflow of the progenitor if its radius exceeded $(40-75)\,R_{\odot}$. Models have been proposed for low-metallicity massive stars that never become red supergiants and indeed reach maximum radii of $\lesssim 40\,R_{\odot}$. These include, for example, some early models for the progenitor of SN 1987A, which exploded at $R\sim 40\,R_\odot$ \citep[e.g.,][]{Weiss1989}. These models do not, however, match the observed distribution of evolved stars in the HR diagram at LMC metallicity: more red supergiants are observed there than at solar metallicity, while the models predict the opposite \citep{Podsiadlowski1992}. This is likely a result of weaker winds at low metallicity, which were not accounted for in early calculations.

More recent models predict that $(10-20)\,M_{\odot}$ stars at low metallicity do expand to become red supergiants, but only do so during their late-stage evolution, during core He burning rather than in the Hertzsprung gap. At $Z=0.01\,Z_\odot$ (lower than \gn), a $14\,M_\odot$ model calculated by \citet{Klencki2020}  spends almost all of its He-burning lifetime with $R<50\,R_{\odot}$, only expanding to $R>100\,R_{\odot}$ in the final $10^5$\,yr before supernova, when the envelope is tenuous and might be easier to unbind. However, similar models at $Z=0.1\,Z_{\odot}$ expand to $R=300\,R_{\odot}$ soon after beginning He burning. We conclude that \gn's low metallicity is unlikely to have prevented the NS progenitor from overflowing its Roche lobe. 

\subsubsection{Expected helium star progenitor}
\label{sec:progenitor}
Models predict that mass transfer from red supergiants proceeds until all but a few tenths of a solar mass of hydrogen have been stripped, and that most of the residual hydrogen will be removed by winds \citep[e.g.,][]{Yoon2017, Gilkis2019}. Given the presence of the luminous star, it is likely that the progenitor of the NS was stripped by binary mass transfer and terminated its evolution as a hydrogen-poor supernova. 

The mapping between He star mass and compact remnant mass is expected to be complex and non-monotonic, and is complicated further by mass transfer in a binary \citep[e.g.,][]{Laplace2021}. In one recent set of models, \citet{Vartanyan2021} predict that NSs with masses near $1.9\,M_{\odot}$ are produced from stripped He stars with masses of $(5-7)\,M_{\odot}$. These form from stars with initial masses of $(17-20)\,M_{\odot}$ in their solar-metallicity models, and likely from somewhat lower initial masses at lower metallicity. Lower-mass He stars are predicted to produce lower-mass NSs, with $M < 1.6\,M_\odot$. Other models in the literature predict minimum He star masses for massive NSs ($M\gtrsim 1.8\,M_\odot$) of $\sim 3$ to $7\,M_{\odot}$ \citep{Mandel2020, Burrows2020, Antoniadis2022}.
 
\subsubsection{Constraints on mass loss and kicks}
\label{sec:blauw}
We use Monte Carlo simulations to study the combinations of pre-supernova orbits and kick velocities and orientations that could have produced \gn. This approach follows \citet{Tauris2017} and \citet{Larsen2024}; for a general review on the effects of SNe in binaries, see \citet{Tauris2023}. In brief, we simulate a large number ($N=10^8$) of orbits and kicks and use the formalism from \citet{Brandt1995} to predict the post-SN parameters. We assume that prior to the formation of the NS, the orbit was circular and the NS progenitor was a He star of mass $M_{\rm He}$. This object's mass instantaneously drops to $M_{\rm NS}=1.90\,M_{\odot}$ when the NS forms. In addition, the NS receives a kick of velocity $v_{\rm kick}$ in a random direction. We fix the luminous star mass to $0.79\,M_{\odot}$ and the NS mass to $1.9\,M_{\odot}$. Finally, we consider the initial parameters of the simulations for which the final period and eccentricity are close to the observed values for \gn: $P_{\rm orb, final} = 700-760$\,d, and $e_{\rm final} = 0.11-0.13$. Narrowing these ranges reduces the number of surviving Monte Carlo samples but has no significant effect on their distribution.



We begin with a uniform distribution of pre-SN orbital periods between 0 and 1000\,d. We simulate kicks assuming velocities distributed uniformly between 0 and 100\,$\rm km\,s^{-1}$, and take He star masses distributed uniformly between $1.9$ and $5.0\,M_\odot$. This is not a population synthesis simulation -- pre-SN orbits are unlikely to be uniformly distributed in all parameters -- but should be viewed as an exploration of what combinations of orbits, kicks, and mass loss could have produced the observed orbit.

The results of this experiment are shown in Figure~\ref{fig:kickplots}. Pre-SN periods ranging from $\sim 400$ to $\sim 900$ d can produce the observed orbit for suitable combinations of mass loss and kicks. Most of the accepted kicks have $\phi_{\rm kick}\approx 180$ deg, meaning that the kick points in the opposite direction of the He star's motion at the time of the SN. The median $v_{\rm kick}$ for SNe that match the observed orbit is $7.9\,\rm km\,s^{-1}$, and 90\% have $v_{\rm kick} < 32\,\rm km\,s^{-1}$. The median He star mass is $2.5\,M_{\odot}$, and 90\% of accepted simulations have $M_{\rm He\,star} < 4.1\,M_{\odot}$.

Although the parameters of \gns can be produced with a significant kick (up to $\sim 80\,\rm km\,s^{-1}$) for fine-tuned orientations and He star masses, {\it most} NS + MS star binaries are expected to end up in more eccentric orbits in this case. This is illustrated in  Figure~\ref{fig:kick}, which shows the predicted median and middle $68\%$ range of the eccentricities as a function of He star mass and kick speed for a pre-SN period of 730 d. Only binaries that remain bound are considered. In the case of no kick (top panel), the eccentricity depends only on the mass lost during the NS's formation: $e=\left(M_{{\rm He}}-M_{{\rm NS}}\right)/\left(M_{{\rm NS}}+M_{\star}\right)$ \citep{Blaauw1961, Hills1983}. In this case, the He star progenitor could not have had a mass larger than $2.2\,M_{\odot}$, corresponding to a total mass loss of $0.3\,M_{\odot}$ during the SNe. Given that neutrinos alone are expected to remove $0.2-0.3\,M_{\odot}$ during the formation of a $1.90\,M_{\odot}$ NS \citep{Lattimer2001}, this leaves little room for any other mass loss in the absence of a kick. 
 
The lower panels of Figure~\ref{fig:kick} show predictions of simulations in which the kick velocity is increased to 5, 10, and $\rm 20\,km\,s^{-1}$. This generally increases the predicted eccentricity, such that in simulations with $v_{\rm kick} > 5\,\rm km\,s^{-1}$, most binaries end up with a higher eccentricity than is observed for \gn, regardless of the progenitor He star's mass. 

Although larger kicks tend to produce higher eccentricities and unbound orbits, there is always a small probability that a fortuitously aimed kick with velocity comparable to the luminous companion's pre-supernova orbital velocity can result in a bound, low-eccentricity orbit. This essentially requires the kick to be oriented such that the NS chases the fleeing companion, and so such kicks result in a net systemic velocity to the binary.

Figure~\ref{fig:low_ecc_prob} explores the probability of such a fine-tuned kick for a variety of He star masses and kick velocities. For each kick velocity, there is a He star mass that maximizes the probability of a bound, low-eccentricity orbit; this is approximately the mass for which the \citet{Blaauw1961} kick velocity is equal to the NS's natal kick. The probabilities in question are all rather low except when the kick is small and the He star has low mass. For $v_{\rm kick}=10\,\rm km\,s^{-1}$, the maximum probability of producing a system like \gns occurs for $M_{\rm He\,star} \approx 3.5\,M_{\odot}$ and is about 8\%. For $v_{\rm kick}=15\,\rm km\,s^{-1}$, the maximum probability falls to $\approx 4\%$ at $M_{\rm He\,star}\approx 5\,M_{\odot}$.

Light curve modeling of stripped-envelope supernovae suggests typical ejecta masses of $(1-3)\,M_{\odot}$ \citep{Lyman2016, Taddia2018}, which for \gns would correspond to He star masses of $\sim (3-5)\,M_{\odot}$. This is within the range of plausible progenitors we infer for kick velocities of $v_{\rm kick}=5-20\,{\rm km\,s^{-1}}$, though only a minority of kick orientations would result in an orbit as circular as \gns under these conditions. It remains to be seen whether other binaries like \gns exist with higher eccentricities.

\subsection{Comparison to other NSs}
\begin{figure*}
    \centering
    \includegraphics[width=\textwidth]{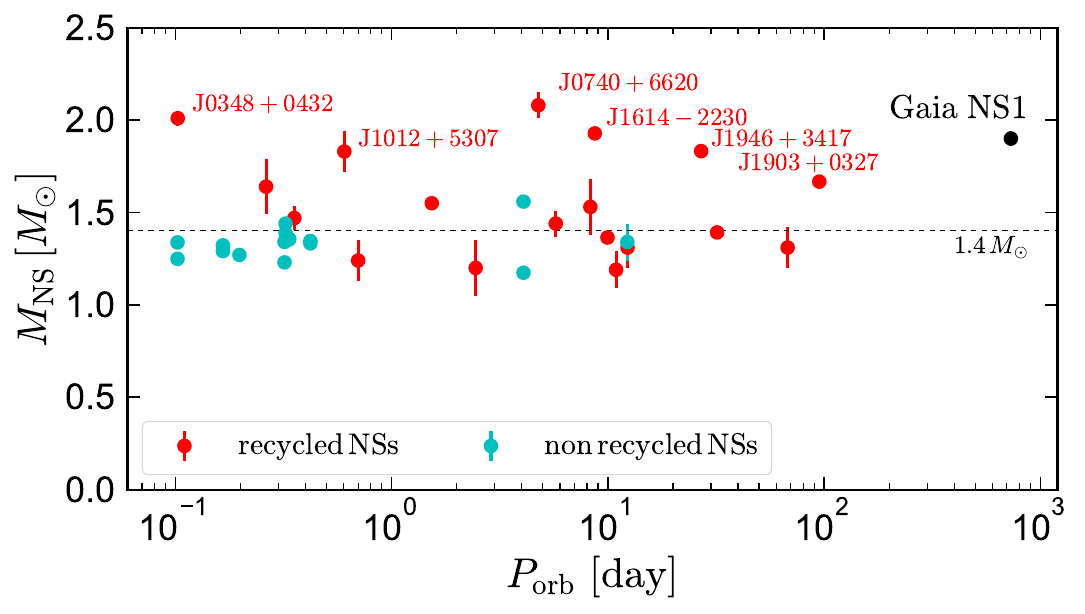}
    \caption{Masses and periods of NSs with well-characterized orbits and mass measurements compared to \gn. Cyan points show radio pulsars that are non-recycled neutron stars (mainly in double NS binaries and a few NS + massive WD binaries), while red points show recycled NSs (millisecond pulsars with low-mass WD companions). Compared to these systems, \gns hosts one of the most massive NSs in the widest orbit. Its parameters are particularly unusual given that the NS presumably has not accreted from the luminous star. This suggests that the NS was born massive (it is not recycled, and is most similar to the cyan points). }
    \label{fig:masses}
\end{figure*}

Figure~\ref{fig:masses} compares our constraints on the mass and orbital period of \gns to known NSs in binaries with well-measured masses, the parameters of which we take from \citet{Ozel2016} and \citet{Fonseca2021}. Cyan points show non-recycled neutron stars; i.e., systems with NS spin periods of order 1\,s and magnetic fields of order $10^{12}$\,G, which have (presumably) not been spun up by accretion. Most of these systems are NS+NS binaries with eccentric orbits and high-precision mass constraints from post-Keplerian effects. Red points show recycled pulsars, most of which have spin periods of a few milliseconds and low-mass WD companions. NSs in spider binaries \citep[e.g][]{Romani2022} are not included.

The recycled NSs generally have larger mass uncertainties because the orbits are circular (leading to fewer measurable post-Keplerian parameters) and the constraints in some cases depend on spectroscopic estimates of the mass of the WD companions. Most of the highest-mass NSs are recycled, meaning that they have accreted significant material and their present-day mass may be significantly higher than their birth mass. Most of the non-recycled NSs have masses near $1.3\,M_{\odot}$, with only two having masses above the Chandrasekhar mass. 

\gns is the fourth most massive object in Figure~\ref{fig:masses}, with only the recycled pulsars J0740+6620, J0348+0432, and J1614-2230 having higher masses. If the dark object is indeed a single NS, there is no plausible scenario in which it gained mass since its formation, so it would be one of the strongest  known cases for a NS being born massive. That being said, J0348+0432 is only mildly recycled, implying that it may have been born with a mass near $2.0\,M_{\odot}$ \citep{Antoniadis2013}, and \citet{Tauris2011} have argued on evolutionary grounds that J1614-2230 was born with a mass of at least $1.6\,M_{\odot}$, and probably higher. There are also some reports of young NSs with relatively high masses in high-mass X-ray binaries \citep[e.g.,][]{Barziv2001, Falanga2015}; these systems are not included in Figure~\ref{fig:masses} because their masses generally carry higher uncertainties. 

\gns is also the longest-period NS in a binary with a precise mass estimate. A few pulsars in longer-period orbits are known \citep[e.g.,][]{Wang2004, vanderWateren2023}, but without precise mass measurements. When the luminous star ascends the giant branch, it will begin transferring mass to the companion, first by winds, and then by stable Roche lobe overflow. This likely makes Gaia NS1 the first known progenitor of a symbiotic X-ray binary \citep[a NS accreting from a red giant; e.g.][]{Yungelson2019}.

\section{Conclusion}
\label{sec:conclusion}

We have presented discovery of \gn, a relatively nearby ($d\approx 735$\,pc) binary containing a low-metallicity turn-off star ($M_\star \approx 0.79\,M_{\odot}$) and an unseen massive companion that we suspect is a neutron star (NS) in a 2-yr orbit. Joint modeling of the system's {\it Gaia} astrometry and our follow-up RVs spanning most of an orbital period allows us to constrain the system's parameters with few-percent precision (Tables~\ref{tab:system}). Our main conclusions are as follows: 

\begin{enumerate}
    \item {\it Properties of the luminous companion}: The luminous companion is a halo star near the main sequence turnoff (Figure~\ref{fig:inc_fig}). Its low metallicity and $\alpha-$enhanced abundance pattern ($\rm [Fe/H]=-1.23$; $[\rm \alpha/Fe]=+0.20$), high space velocity ($v_{\rm pec}\approx 300\,\rm km\,s^{-1}$) and ancient age ($\tau \gtrsim 12$\,Gyr) all suggest that it is a halo star, probably formed in a since-accreted satellite of the Milky Way. 
    
    The star's abundance pattern (Table~\ref{tab:abund}) is in most respects unremarkable. However, the star is highly enriched in lithium, with a surface abundance higher than any of the $1000+$ stars with similar parameters observed by the GALAH survey (Figure~\ref{fig:galah_lithium}). This lithium enhancement is almost certainly related to the companion, but exactly how is unclear. Unexplained lithium enhancement has been found in the companions to several BH and NS X-ray binaries. Lithium is also produced in massive AGB stars, suggesting possible pollution from the progenitor of a WD or a NS formed through an electron-capture supernova. However, these scenarios are expected to produce a compact object below the Chandrasekhar mass and are likely only viable for \gns if the companion is a tight binary. 
    
    The star's parameters and evolutionary state imply a mass of $0.79\pm 0.01\,M_{\odot}$ when interpreted with single-star evolutionary models. We expect the mass inferred in this way to be reliable because any mass transfer would have occurred many Gyr ago in all the evolutionary scenarios we consider. The star's broadband spectral energy distribution (Figure~\ref{fig:inc_fig}) and high-resolution spectra (Figure~\ref{fig:galah}) are well-matched by single-star models. Even if we consider the (remote) possibility that the distance could be significantly different from our fiducial assumptions, the star's temperature and metallicity imply a minimum mass of $0.75\,M_{\odot}$. We consider and reject evolutionary scenarios in which the star is a low-mass stripped object, as these do not produce stars with temperatures and surface gravities near the observed values (Section~\ref{sec:lower_mass}). 
    
    \item {\it Orbit and companion mass}: Our RV follow-up covers the full dynamic range of the orbit and thus tightly constrains its parameters. Joint fitting of the astrometry and RVs constrains the companion mass to $M_2=1.90\pm 0.01\,M_{\odot}$ when the {\it Gaia} astrometric uncertainties are taken at face value (Table~\ref{tab:system}). The RVs evolve in a way that is basically consistent with the predictions of the pure-astrometry solution (Figure~\ref{fig:rvfig}) and there are no significant deviations from a Keplerian orbit. However, the observed RV variability amplitude is $\sim 3\,{\rm km\,s^{-1}}$ smaller than the median value predicted by the astrometric solution, implying a smaller physical scale for the orbit. For this reason, the best-fit companion mass inferred from our joint RV+astrometry fit is $\sim 2\sigma$ lower than the value inferred from astrometry alone. 

    There is moderate tension between the parameter estimates from the astrometric orbit alone and our joint astrometry + RV fit (Figure~\ref{fig:corner}). The most discrepant parameter between the two sets of constraints is the parallax, for which the {\it Gaia}-only constraint is $\varpi=1.24\pm 0.04$ mas but the constraint from joint fitting of the astrometry and RVs is $\varpi=1.36\pm 0.01$\,mas. We ascribe this tension to underestimated uncertainties in the astrometric solution, as was also found for Gaia BH1 \citep{Chakrabarti2023, Nagarajan2023}. We therefore inflated all the astrometric uncertainties by a factor of 3 and repeated the joint RV + astrometry fit. In this case, constraints between the astrometry-only and astrometry+RVs fit are consistent at the $1\sigma$ level, and the companion mass is constrained to $M_2 = 1.90\pm 0.04\,M_{\odot}$.

    The orbit is nearly circular, with an eccentricity $e=0.122\pm 0.003$. The inferred companion mass is only weakly dependent on the {\it Gaia} astrometry, because the orbit is relatively edge-on ($i=69$ deg; Figure~\ref{fig:inc_fig}). If the orbit were edge-on ($i=90$ deg), the RVs and luminous star mass would imply a companion mass of $M_2=1.67\,M_{\odot}$. 
    
    \item {\it Nature of the companion}: We have not detected any electromagnetic radiation from the companion, so our speculation about its nature is informed only by our constraints on its mass and considerations of possible formation histories. Its mass is clearly too high to be a white dwarf, and likely too low for it to be an astrophysical BH. Luminous companions are ruled out because they would imply even higher companion masses (Figure~\ref{fig:scenarios}) and RV amplitudes that are inconsistent with our observations; indeed, joint fitting of the astrometry and RVs sets an upper limit on the flux ratio of $\epsilon < 0.02$, even with the astrometric uncertainties inflated. A single, relatively massive NS thus appears to be the simplest possible explanation.  

    The only plausible alternative is that the dark companion is a tight binary containing two massive WDs, or a WD and a NS. This scenario is consistent with the data but difficult to explain with evolutionary models (see Section~\ref{sec:companion_nature} for discussion). Models in which the inner binary remained tight over its whole evolution fail to form sufficiently massive WDs, and to explain the observed lithium pollution of the luminous star. Models in which both inner and outer companions to a massive star went through simultaneous common envelope events provide a potential solution but are very uncertain. 
    
    \item {\it Formation history}: The system's formation history is also difficult to explain if the dark object is a single high-mass NS. The low eccentricity makes dynamical formation channels (i.e., exchange interactions in a dense cluster) unlikely. Assuming the system formed from a primordial binary, the progenitor of the NS likely had an initial mass of $10-20\,M_{\odot}$. A star of this mass is expected to become a red supergiant that would have overflowed its Roche lobe in the current orbit. Because the donor-to-accretor mass ratio would be highly unequal in this scenario, mass transfer is expected to be unstable. However, unstable mass transfer (i.e., common envelope evolution) is predicted to often result in a much tighter orbit than observed. These considerations suggest that (a) mass transfer was stable despite a highly equal mass ratio, (b) common envelope ejection was very efficient, or (c) the system did not form via isolated binary evolution. This is essentially the same ``problem'' that has been pointed out for isolated binary evolution channels forming Gaia BH1 and BH2.

    \item {\it Constraints on natal kicks}: The wide and near-circular orbit of \gns today places strong constraints on mass loss and kicks during the NS's formation (Figures ~\ref{fig:kickplots},~\ref{fig:kick} and~\ref{fig:low_ecc_prob}). The simplest explanation for the observed orbit is a weak natal kick ($v_{\rm kick}\lesssim 20\,{\rm km\,s^{-1}}$) and little ejected mass ($M_{\rm He\,star}\lesssim 3\,M_{\odot}$, corresponding to an ejecta mass of $\lesssim 1\,M_{\odot}$.) Stronger kicks are possible if coupled with more mass loss, but only for fine-tuned kick orientations. 
\end{enumerate}


Like Gaia BH1 and BH2, \gns is an unexpected discovery. The system's wide orbit and low-mass luminous star make it difficult to form through the standard evolutionary channels invoked for the formation of low- and high-mass X-ray binaries and millisecond pulsars. On the other hand, a few low-mass red giant + NS binaries are known in symbiotic binaries with orbital periods comparable to \gns \citep[e.g.,][]{Hinkle2006}. Such systems must have formed from wide MS + NS binaries like \gns and are X-ray bright only for the short period in which the companion is a giant. There is thus little doubt that wide low-mass MS + NS binaries exist, and we suspect that \gns is the first  clear example of one to be identified.
 
\section*{acknowledgments}
We thank the referee for a constructive report, and Jim Fuller, Eliot Quataert, and Paulo Freire for useful discussions. We are grateful to Yuri Beletsky, Sam Kim, Angela Hempel, Régis Lachaume, Gil Esquerdo, Perry Berlind, Mike Calkins, and Casey Lam for observing help.  This research was supported by NSF grant AST-2307232. HWR acknowledges the European Research Council for the ERC Advanced Grant [101054731]. This research benefited from discussions in the ZTF Theory Network, funded in part by the Gordon and Betty Moore Foundation through Grant GBMF5076, and from collaboration at the ``Renaissance of Stellar Black-Hole Detections in The Local Group'' workshop hosted at the Lorentz Center in June, 2023.

This research made use of pystrometry, an open source Python package for astrometry timeseries analysis \citep[][]{Sahlmann2019}. This work made use of Astropy,\footnote{http://www.astropy.org} a community-developed core Python package and an ecosystem of tools and resources for astronomy \citep{AstropyCollaboration2022}.

This work has made use of data from the European Space Agency (ESA) mission
{\it Gaia} (\url{https://www.cosmos.esa.int/gaia}), processed by the {\it Gaia}
Data Processing and Analysis Consortium (DPAC,
\url{https://www.cosmos.esa.int/web/gaia/dpac/consortium}). Funding for the DPAC
has been provided by national institutions, in particular the institutions
participating in the {\it Gaia} Multilateral Agreement.

This paper includes data gathered with the 6.5 meter Magellan Telescopes located at Las Campanas Observatory, Chile. 

\newpage

\newpage

\appendix

\section{Joint RVs + astrometry fit with inflated uncertainties}
\label{sec:appendix}

Figure~\ref{fig:corner} compares the astrometry-only and astrometry+RVs constraints when the {\it Gaia} uncertainties are inflated by a factor of 3. In this case, the two sets of constraints are consistent within 1$\sigma$ for all parameters. For most parameters, they are consistent to within significantly less than $1\sigma$, suggesting that inflating the {\it Gaia} uncertainties by a factor of 3 is conservative.

\begin{figure*}
    \centering
    \includegraphics[width=0.9\textwidth]{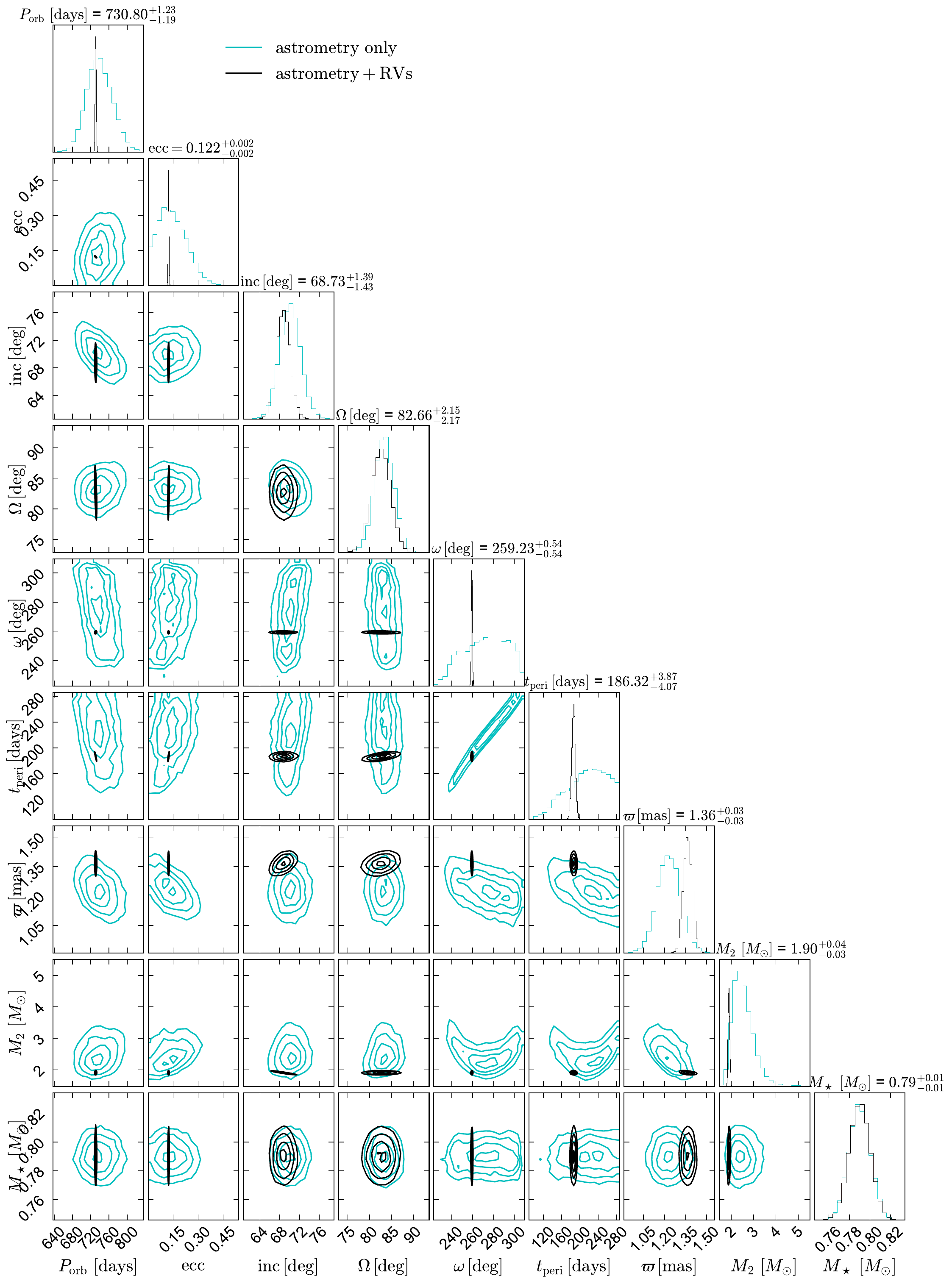}
    \caption{Similar to Figure~\ref{fig:corner}, but for the case with the {\it Gaia} uncertainties inflated by a factor of 3. Constraints on all parameters are then consistent within 1$\sigma$ between the astrometry only and astrometry+RVs fit.}
    \label{fig:corner_inflate}
\end{figure*}

\end{document}